# Rigorous modal analysis of plasmonic nanoresonators

Authors: Wei Yan*, Rémi Faggiani, and Philippe Lalanne*

Affiliations:

Laboratoire Photonique, Numérique et Nanosciences (LP2N), IOGS – Univ. Bordeaux – CNRS, 33400 Talence cedex, France.

*Correspondence to: philippe.lalanne@institutoptique.fr, wei.yan@institutoptique.fr

**Abstract.** The specificity of modal-expansion formalisms is their capabilities to model the physical properties in the natural resonance-state basis of the system in question, leading to a transparent interpretation of the numerical results. In electromagnetism, modal-expansion formalisms are routinely used for optical waveguides. In contrast, they are much less mature for analyzing open non-Hermitian systems, such as micro and nanoresonators. Here, by accounting for material dispersion with auxiliary fields, we considerably extend the capabilities of these formalisms, in terms of computational effectiveness, number of states handled and range of validity. We implement an efficient finite element solver to compute the resonance states, and derive new closed-form expressions of the modal excitation coefficients for reconstructing the scattered fields. Together, these two achievements allow us to perform rigorous modal analysis of complicated plasmonic resonators, being not limited to a few resonance states, with straightforward physical interpretations and remarkable computation speeds. We particularly show that, when the number of states retained in the expansion increases, convergence towards accurate predictions is achieved, offering a solid theoretical foundation for analyzing important issues, e.g. Fano interference, quenching, coupling with the continuum, which are critical in nanophotonic research.

## I. Introduction

The control of light at the nanoscale is ultimately limited by our capability to engineer electromagnetic near-fields with several nanoresonances, enable energy transfers between them, and model how every individual state precisely interfere to create new resonant states that overlap in space and energy. Optical nanoresonators, be they plasmonic, photonic or both, offer a unique route to enhance and localize the electromagnetic energy at wavelength or subwavelength scales. They now play a leading role in many areas in nanophotonics, from quantum information processing to ultrasensitive biosensing, nonlinear optics, and various optical metasurfaces.

Classically, the resonant interaction of light with optical resonances is modeled via continuum scattering theory with Maxwell solvers operating in the time or real frequency domains. As they do not explicitly compute the resonance states, the numerical predictions are not always easy to interpret - the black-box sensation often experienced by users. They additionally require many computations that must be repeated for every individual frequency in the frequency-domain, or in the time-domain for every instance of the driving field, e.g. the pulse duration, polarization, incidence angle [1,2]. In sharp contrast, the present formalism operates at complex frequencies $\widetilde{\omega}$ set by the natural states of the resonator, also called quasinormal modes (QNMs). It offers two decisive advantages. It brings the physics of the resonant states at the heart of the analysis, thus removing the black-box sensation. It also allows to model properties that span over a frequency range $\sim \text{Im}(\widetilde{\omega})$ around a central frequency $\text{Re}(\widetilde{\omega})$ with a high degree of analyticity, which often offer unprecedented modeling capabilities and computational performance. The formalism belongs to the general category of modal formalisms, and as such, is expected to bring a valuable input to resonance optics, comparable to that previously brought by the guided mode theory to integrated optics design [3-4].

The spectral representation of waves in resonant systems as a superposition of QNMs has a venerable story [5-9]. In electromagnetism, pioneer works initially focused on simple geometries, e.g. 1D and 3D-spherical resonators in uniform backgrounds, for which the completeness of QNM-expansions was established [10-11]. Important follow-up results were concerned by extensions of these founding works towards arbitrary resonant geometries. They include a mathematically-sound normalization and definition of the mode volume [12-15], the derivation of orthogonality relation for QNMs of non-dispersive resonators [10,12,16], the elaboration of perturbation methods for calculating the QNMs of a resonator from the knowledge of the QNMs of another unperturbed resonator [16-19], attempts to implement numerically stable methods to compute and normalize QNMs [20-25]. For a recent review, see [15].

Many ingredients towards a complete modal theory of nanoresonators are now available, but they are scarcely presented and often restricted to peculiar geometries. Hereafter, we extend these works to propose a comprehensive theoretical framework for the most general case of 3D resonators with arbitrary shapes and materials, possibly in non-uniform backgrounds. The generality and effectiveness of the framework are validated by implementing an effective QNM-expansion software. Of particular importance in the present context is the successful generalization of the auxiliary-field method, originally proposed for simulating dispersive media with finite difference time-domain simulations [26] and computing the band diagram of dispersive crystals [27], to compute the QNMs of open resonators with finite element methods (FEMs). This allows us to successfully implement a QNM solver that efficiently computes the eigenstates of plasmonic resonators. The achieved precision is much better than those offered by finite difference approaches [21,27], especially for the usual cases of metallic nanoresonators with curved shapes. On the theoretical side, another important consequence of the auxiliary-field method is a net physical interpretation of temporal dispersion, which lead us to derive orthogonality relations in the augmented formulation for resonances made of dispersive media. Such a derivation that was not possible in earlier works with unspecified dispersion relation [12,15] leads to the important proposition of closed-form expressions for the eigenstate excitation coefficients.

We believe that the joint effort in numerics and theory greatly expands the capabilities of analyzing electromagnetic resonance in nanophotonics, offering increased physical insight and improved computational speed. This brings us closer to a comprehensive modal theory of optical resonances, the analogue of the optical waveguide theory for nanoresonators.

## II. Augmented-field QNM-formulation

The QNMs of localized resonators, made of dispersive or non-dispersive media or both, are defined as the time harmonic solutions to the source-free Maxwell's equations

$$\begin{bmatrix} 0 & -i\mu_0^{-1}\boldsymbol{\nabla}\times \\ i\varepsilon(\widetilde{\omega}_m)^{-1}\boldsymbol{\nabla}\times & 0 \end{bmatrix} \begin{bmatrix} \widetilde{\mathbf{H}}_m \\ \widetilde{\mathbf{E}}_m \end{bmatrix} = \widetilde{\omega}_m \begin{bmatrix} \widetilde{\mathbf{H}}_m \\ \widetilde{\mathbf{E}}_m \end{bmatrix}, \qquad (1)$$

with $\varepsilon(\widetilde{\omega}_m)$ the permittivity and $\mu_0$ the vacuum permeability, an $\exp(-i\widetilde{\omega}_m t)$ dependence being assumed for time harmonic fields. The QNMs defined by their electric and magnetic field vectors, $\widetilde{\mathbf{E}}_m$ and $\widetilde{\mathbf{H}}_m$, satisfy the outgoing-wave conditions, have complex frequency $\widetilde{\omega}_m$ with quality factor $Q_m = -\frac{1}{2}\mathrm{Re}(\widetilde{\omega}_m)/\mathrm{Im}(\widetilde{\omega}_m)$, and grow exponentially at large distance from the structure [15].

For dispersive materials, a difficulty arises as the eigenproblem of Eq. (1) no longer defines a standard linear eigenproblem. The nonlinearity, which essentially arises from hidden variables that are eliminated in the constitutive relations, has a prescribed nature. At optical frequencies, most material properties can be modeled with a standard $N$-pole Lorentz-Drude relationship [27,29], $\varepsilon(\omega) = \varepsilon_\infty - \varepsilon_\infty \sum_i \omega_{p,i}^2 (\omega^2 - \omega_{0,i}^2 + i\omega\gamma_i)^{-1}$, with notations for the plasma frequencies $\omega_{p,i}$, the damping coefficients $\gamma_i$ and the resonant frequencies $\omega_{0,i}$. In metals for instance, the intraband transition gives rise to free-electron behavior characterized by a Drude pole ($\omega_{0,i} = 0$), whereas interband transitions are faithfully represented by Lorentz poles ($\omega_{0,i} \neq 0$) [29]. As already noted in [21,26-27], the nonlinear eigenproblem for Lorentz-Drude materials can be cast into a linear one by reintroducing hidden auxiliary fields, such as the polarization $\mathbf{P}_i = -\varepsilon_\infty \omega_{p,i}^2 (\omega^2 - \omega_{0,i}^2 + i\omega\gamma_i)^{-1} \mathbf{E}$ and the current $\mathbf{J}_i = -i\omega \mathbf{P}_i$. Considering a single Lorentz-pole permittivity to simplify the notations and denoting by $\widetilde{\boldsymbol{\Psi}}_m = [\widetilde{\mathbf{H}}_m, \widetilde{\mathbf{E}}_m, \widetilde{\mathbf{P}}_m, \widetilde{\mathbf{J}}_m]^\mathbf{T}$ the augmented eigenvector, Eq. (1) is reformulated in a linear form

$$\widehat{\mathbf{H}}\widetilde{\boldsymbol{\Psi}}_m = \begin{bmatrix} 0 & -i\mu_0^{-1}\boldsymbol{\nabla}\times & 0 & 0 \\ i\,\varepsilon_\infty^{-1}\boldsymbol{\nabla}\times & 0 & 0 & -i\varepsilon_\infty^{-1} \\ 0 & 0 & 0 & i \\ 0 & i\omega_p^2\varepsilon_\infty & -i\omega_0^2 & -i\gamma \end{bmatrix} \widetilde{\boldsymbol{\Psi}}_m = \widetilde{\omega}_m \widetilde{\boldsymbol{\Psi}}_m, \qquad (2)$$

in the Lorentz-material subspace and takes the usual form without auxiliary fields elsewhere.

**A. Discretized versus continuous operators: QNM and PML-modes**

We should bear in mind that the following numerical and theoretical results are all obtained for a slightly different version of the original physical problem, in which the original open space is replaced by a finite space bounded with perfectly-matched layers (PMLs) implementing the outgoing-wave conditions. Thus the continuous operator (Eq. 2), originally defined on an unbounded space, after the PML mapping and numerical discretization, is replaced by an analytically-continued, discretized operator (a finite-dimensional matrix) defined on a finite mapped space, with new permittivity and permeability distributions that accommodate the PMLs. With classical frequency-domain Maxwell solvers, the matrix $\widehat{\mathbf{H}} - \omega\widehat{\mathbf{I}}$ ($\widehat{\mathbf{I}}$ being the identity matrix) is inverted for each frequency $\omega$ in order to compute the electromagnetic response to an external source. Since the discretization and the PMLs are effective only within a finite frequency interval (or domain $\mathcal{F}$ of the complex-$\omega$ plane), the electromagnetic solutions of the analytically-continued operator are accurate only for frequencies within $\mathcal{F}$.

Here, instead of inverting $\widehat{\mathbf{H}}$ for every frequency, we consider the discrete spectrum of $\widehat{\mathbf{H}}$, and reconstruct the electromagnetic solution *analytically* from the spectral decomposition. Since $\widehat{\mathbf{H}}$ is linear, using the matrix inversion scheme or the spectral decomposition approach leads to the same results (up to numerical uncertainty), i.e. the spectral decomposition approach provides exactly the same solutions as the inversion approach, if the whole set of eigenstates is considered. In particular, we get the same faithful solutions within $\mathcal{F}$, and even more importantly, the QNMs of the original continuous operator with $\widetilde{\omega}_m \in \mathcal{F}$ are accurately recovered in the spectrum of $\widehat{\mathbf{H}}$ and the physics – the natural resonances – is preserved. Advantageously, the new mapped eigenstates do not grow exponentially, become square-integrable and can be normalized [12]. In addition, complicated theoretical issues on the completeness of QNM expansions for open spaces are avoided. Indeed, the eigenstates of $\widehat{\mathbf{H}}$ form a complete set over the whole mapped space, in contrast with the QNMs of the open system that form, at best, a complete set only "inside the resonator" [10]. This has important consequences for instance for plasmonic dimers like bowtie antennas, since by including the PML-modes, the expansion becomes complete even outside the metal, for instance in the dielectric gap where interesting phenomena occur due to the strong field enhancement.

The eigenstates of the mapped operator are composed of two sets of modes, a subset of the QNMs of the original continuous operator and a set of numerical modes, called hereafter PML-modes [28,15], which depend on the PML parameters and form a complete basis with the QNM subset. The PML-modes can be further classified into two subclasses. The first class is fed with distorted remnants of the original QNMs with $\widetilde{\omega}_m \notin \mathcal{F}$, which cannot be recovered with the discretized operator. The second class originates from the continuum of background modes [28,30]. For thick PMLs, they are Fabry-Perot modes of closed resonators mainly formed by the PMLs, and are composed of ingoing and outgoing waves in the PMLs, unlike QNMs. For uniform backgrounds, this second class of eigenstates is easily detected as it forms a tilted straight line in the complex plane, as in Fig. 1a, resulting from a rotation by the complex stretching of the PMLs of the original continuum of plane waves that lies on the real-frequency axis [28,30]. For non-uniform backgrounds, the locations of the PML-mode and QNMs in the complex plane are entangled, and disentangling may be challenging as illustrated with the last numerical example of the article. Supplementary Section 3.4 provides more details on how one may distinguish QNMs and PML-modes in this case. However the distinction is not always required, as in general, both QNMs and PML-modes have to be considered in the expansion for numerical accuracy.

## B. QNM eigensolver implementation

Based on Eq. (2), we have developed a QNM solver using the COMSOL Multiphysics environment [31]. For that purpose, Eq. (2) is first transformed into a standard quadratic eigenvalue problem

$$\widehat{\mathbf{K}} \begin{bmatrix} \widetilde{\mathbf{E}}_m \\ \widetilde{\mathbf{P}}_m \end{bmatrix} + \widetilde{\omega}_m \widehat{\mathbf{C}} \begin{bmatrix} \widetilde{\mathbf{E}}_m \\ \widetilde{\mathbf{P}}_m \end{bmatrix} + \widetilde{\omega}_m^2 \widehat{\mathbf{M}} \begin{bmatrix} \widetilde{\mathbf{E}}_m \\ \widetilde{\mathbf{P}}_m \end{bmatrix} = 0, \tag{3}$$

for which stable and efficient algorithms exist [32]. In Eq. (3), $\widehat{\mathbf{K}} = \begin{bmatrix} \mathbf{\nabla} \times \mu_0^{-1} \mathbf{\nabla} \times & 0 \\ \varepsilon_\infty & -\omega_0^2 \end{bmatrix}$ is the stiffness matrix, $\widehat{\mathbf{C}} = \begin{bmatrix} 0 & 0 \\ 0 & i\gamma \end{bmatrix}$ the damping matrix, and $\widehat{\mathbf{M}} = \begin{bmatrix} -\varepsilon_\infty & -1 \\ 0 & 1 \end{bmatrix}$ the mass matrix. Then the coupled system of partial differential equations is converted into its equivalent weak formulations that are entered directly in the COMSOL environment using the user COMSOL-interface framework for partial differential equations. Finally the discretized equations are solved with the built-in iterative eigensolver of COMSOL .

The solver could be conveniently used to compute the QNMs and PML-modes of arbitrary, 3D plasmonic resonators in complex photonic environments, e.g., resonators on substrates. It takes full advantage of the power of finite-element meshes to accurately model geometries mixing curved shapes, large-scale features, e.g., planar interfaces, and small-scale features, e.g., sharp corners, which ensures superior accuracy and performance compared to previous solvers developed with finite-difference discretization schemes [21,27]. Moreover, since the additional auxiliary fields are restrictively defined in the computational subspaces with dispersive materials only, the matrix-size increase due to the auxiliary fields remains moderate. Details of the weak-formulation and numerical tests of the solver performance can be found in Supplementary Section 4.

To illustrate the QNM-solver capabilities, we consider a silver bowtie antenna in a uniform air background with a permittivity $\varepsilon_b = 1$. Figure 1a shows the energies and decay rates of the eigenstates computed with the solver. QNMs and PML-modes are shown with blue circles and gray squares, respectively. The QNMs can be decomposed into two subsets. "Transverse" QNMs satisfying $\mathbf{\nabla} \cdot \varepsilon(\widetilde{\omega}_m) \widetilde{\mathbf{E}}_m = 0$ and $\mathbf{\nabla} \cdot \mu_0 \widetilde{\mathbf{H}}_m = 0$ are obtained for $\widetilde{\omega}_m \neq 0$ and are directly computed with the solver. This large subset is enriched by longitudinal QNMs, including bulk plasmons with nonzero eigenfrequencies $\widetilde{\omega}_B$ such that $\varepsilon(\widetilde{\omega}_B) = 0$, and static QNMs computed by solving $\mathbf{\nabla} \times \widetilde{\mathbf{E}}_m = 0$ (electric static) or $\nabla \times \widetilde{\mathbf{H}}_m = 0$ (magnetic static); in this case, electric static QNMs have null electric fields inside resonators since $\varepsilon \to \infty$ as $\omega \to 0$. Figure 1b shows the intensity distributions of 4 eigenstates, marked with large squares in (a).

In [33], we provide the COMSOL models and the companion Matlab programs that we have developed for this work. They can be easily reused and updated by drawing new geometries with the COMSOL graphical user interface.

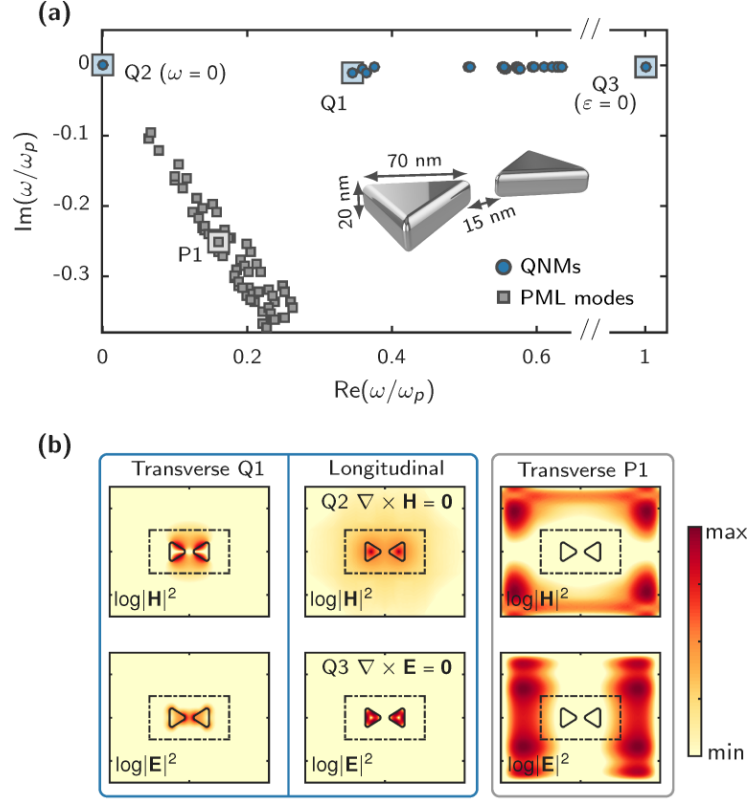

FIG. 1. **QNMs and PML-modes of a silver bowtie nanoantenna.** The bowtie parameters are given in the inset in **(a)** and the rounding radius of the corner is 8 nm. **(a)** Eigenstate energies and decay rates computed with the auxiliary-field solver. The measured silver permittivity [34] is approximated by a realistic Drude model with $\lambda_p = 2\pi c/\omega_p = 138$ nm and $\gamma = 0.0023\omega_p$. Note that the eigenstates always pairwise coexist, $\widetilde{\Psi}_m$ at $\widetilde{\omega}_m$ and $\widetilde{\Psi}_m^*$ at $-\widetilde{\omega}_m^*$ as implied by the Hermitian symmetry $\varepsilon(\widetilde{\omega}_m) = \varepsilon^*(-\widetilde{\omega}_m^*)$, and that we have represented only the eigenstate subset with positive real energies, which are revealed by our non-dispersive PMLs. **(b)** Intensity distributions of the normalized QNMs Q1, Q2, Q3 and the PML-mode P1. We highlight a typical feature of the PML-modes that their field intensities are dominantly located in the PML, the region outside the dash-dotted box.

### III. Reconstruction: Orthogonality relation, modal excitation coefficients

Because it explicitly considers the physical origin of the dispersion in the constitutive relation, the augmented formulation offers a number of advantages compared to nonlinear eigenvalue formulations based on electric and magnetic fields only. For instance, as shown in the Supplementary Section 2.2, it enables an explicit distinction between different forms of the QNM energy at complex frequencies, including the kinetic energy of electrons, Ohmic absorption and radiation leakage, which is not directly available otherwise. It additionally allows the derivation of important relationships, such as an orthogonality relation and a closed-form expression for the modal excitation coefficients used in the spectral representation of waves as superpositions of QNMs. The derivations do not require sophisticated mathematics, and are provided in their entirety in the Supplementary Section 3. Hereafter, we just summarize the main results used for the numerical examples presented in the next Section. These results are valid for the general case of nanoresonators composed of dispersive materials placed in non-uniform backgrounds.

The orthogonality relation between the whole set of PML-modes and QNMs for dispersive nanoresonators directly follows from the unconjugated Lorentz reciprocity theorem applied to the augmented formulation [35], and takes the following form

$$\langle\widetilde{\pmb{\Psi}}_n^*|\widehat{\mathbf{D}}|\widetilde{\pmb{\Psi}}_m\rangle_{V_c} = \iiint_{V_c}\left[\varepsilon_\infty\widetilde{\mathbf{E}}_n\cdot\widetilde{\mathbf{E}}_m - \mu_0\widetilde{\mathbf{H}}_n\cdot\widetilde{\mathbf{H}}_m + \omega_0^2/\varepsilon_\infty\omega_p^2\,\widetilde{\mathbf{P}}_n\cdot\widetilde{\mathbf{P}}_m - 1/\varepsilon_\infty\omega_p^2\,\widetilde{\mathbf{J}}_n\cdot\widetilde{\mathbf{J}}_m\right]d^3\mathbf{r} = \delta_{nm}, \quad (4)$$

with $\delta_{nm} = 1$ if $n = m$ and 0 otherwise, $\widehat{\mathbf{D}} = \mathrm{diag}[-\mu_0, \varepsilon_\infty, \omega_0^2/(\varepsilon_\infty\omega_p^2), -1/(\varepsilon_\infty\omega_p^2)]$. $V_c$ denotes the volume of the entire mapped space, including the PMLs. It is additionally easy to show that the normalization condition $\langle\widetilde{\pmb{\Psi}}_m^*|\widehat{\mathbf{D}}|\widetilde{\pmb{\Psi}}_m\rangle_{V_c} = 1$ can be simply rewritten as $\iiint_{V_c}\left[\frac{\partial\widetilde{\omega}_m\varepsilon(\widetilde{\omega}_m)}{\partial\widetilde{\omega}_m}\widetilde{\mathbf{E}}_m\cdot\widetilde{\mathbf{E}}_m - \mu_0\widetilde{\mathbf{H}}_m\cdot\widetilde{\mathbf{H}}_m\right]d^3\mathbf{r} = 1$, consistently with earlier works [10,12,14,36]. Note that no energy consideration with conjugate scalar products is underpinned by Eq. (4). We emphasize that Eq. (4) differs from the orthogonality relation proposed in [27] for pseudo-periodic auxiliary fields of Bloch modes of plasmonic crystals, owing to the non-Hermitian character of the present QNM eigenvalue problem. Besides, it was not reported in all previous works on QNMs of dispersive resonators with nonlinear eigenvalue problem formulations based on electric and magnetic fields only, for which it was shown that QNMs are not orthogonal, see for instance Eqs. (5)-(6) in [12] and Fig. 8 in [15].

The second important result concerns the spectral representation of the field scattered $\pmb{\Psi}_{\mathrm{sca}}(\mathbf{r},\omega)$ at real frequency $\omega$ by any resonator (reconstruction problem)

$$\pmb{\Psi}_{\mathrm{sca}}(\mathbf{r},\omega) = \sum_m \alpha_m(\omega)\,\widetilde{\pmb{\psi}}_m(\mathbf{r}), \quad (5)$$

for which it is possible to derive a closed-form expression for the excitation modal coefficients [35]

$$\alpha_m(\omega) = \frac{\widetilde{\omega}_m}{\widetilde{\omega}_m - \omega}\langle\widetilde{\mathbf{E}}_m^*|\varepsilon(\widetilde{\omega}_m) - \varepsilon_b|\mathbf{E}_{\mathrm{inc}}(\omega)\rangle_{V_{\mathrm{res}}} + \langle\widetilde{\mathbf{E}}_m^*|\varepsilon_b - \varepsilon_\infty|\mathbf{E}_{\mathrm{inc}}(\omega)\rangle_{V_{\mathrm{res}}}, \quad (6)$$

valid for any near- or far-field illumination. In Eq. (6), $\varepsilon_b$ is the relative permittivity of the medium surrounding the resonator, $\langle\widetilde{\mathbf{E}}_m^*|f(\mathbf{r})|\mathbf{E}_{\mathrm{inc}}(\omega)\rangle_{V_{\mathrm{res}}} = \iiint_{V_{\mathrm{res}}}f(\mathbf{r})\widetilde{\mathbf{E}}_m\cdot\mathbf{E}_{\mathrm{inc}}(\omega)\,d^3\mathbf{r}$ is the usual overlap integral involving the resonance mode and the driving field, $V_{\mathrm{res}}$ being the resonator domain used for the scattered-field formulation and $f(\mathbf{r})$ a weighting function. Equations (5) and (6) provide a high-level of analyticity for the reconstruction step.

As will be evidenced by the following numerical results, the natural resonances, QNMs of the continuous operator are largely recovered in the spectrum of the mapped operator, and thus the set of eigenvectors $\widetilde{\pmb{\psi}}_m(\mathbf{r})$ constitutes a basis of predilection to model the resonator, explicitly highlighting Fano-like interferences between the dominant eigenvectors. In addition, assuming the absence of any accidental degeneracies at exceptional points for instance [37], the eigenvectors form a complete set over the whole mapped space and the QNM expansion provides highly accurate predictions – in [35], we show the following closure relation $\sum_{m=1}^\infty \widetilde{\pmb{\psi}}_m(\mathbf{r}')\widetilde{\pmb{\psi}}_m^{\mathrm{T}}(\mathbf{r})\widehat{\mathbf{D}} = \widehat{\mathbf{I}}\delta(\mathbf{r} - \mathbf{r}')$, $\widehat{\mathbf{I}}$ being an identity matrix and the superscript "T" being the transpose operator. It is important to bear in mind that Eq. (5), like the closure relation, holds for any $\mathbf{r}$ in the whole mapped space, including the PMLs, in sharp contrast with QNM expansions of open systems that are strictly exact only "inside the resonator" [10,11].

Equations (5)-(6) will be repeatedly used for the following numerical results obtained for various nanoresonator geometries and driving illuminations. All eigenvectors do not play the same role in those equations. To realize what the dominant eigenvectors could be, we again consider the spectrum of continuous operators. From rigorous mathematical proof obtained for simple 1D and 3D spherically symmetric resonators geometries, it is widely accepted that the (true) QNM set form a complete basis, provided that the electromagnetic Green's tensor $\mathbf{G}(\omega, \mathbf{r}', \mathbf{r})$ is analytic in the complex-frequency plane excluding the QNM poles and that $\mathbf{r}$ and $\mathbf{r}'$ are inside the resonator [10,11]. This implies that for simple geometries, such as 3D resonators in uniform dielectric backgrounds, the PML-modes of the mapped operator are expected to play a non-dominant role in the expansion of Eq. (5). Thus, provided that the PML mapping preserves the dominant QNMs of the open space, accurate predictions are expected for the scattered field with only QNMs, only for simple geometries. This result will be illustrated by numerical results hereafter.

## IV. Numerical results and validation

With a few selected examples, we test the theoretical results with what is happening on the ground and assess the modeling capabilities of the QNM-expansion formalism, either for clarifying the physics of resonant systems based on the interference of a few modes or for accurately analyzing complex situations involving many modes of different nature. Other pertaining tests on the computational speed, the exactness or the convergence as a function of the number of eigenstates retained in the expansion of Eq. (5) are outlined in the Supplementary Section 5.

### A. Spectral and temporal analysis of antenna in uniform backgrounds

The first example concerns a bowtie antenna in air. The first motivation for studying this relatively simple geometry is to illustrate the importance of QNMs, comparatively to PML-modes, for resonators with uniform backgrounds, by showing that the convergence rate is dominantly driven by QNMs in this case and that PML-modes convey minor additional information that is required only if a high accuracy is desired. Another motivation is to lay new foundations for modal-expansions in the temporal domain, which has received only little attention in the literature to our knowledge [38]. It is yet another motivation to show that the modal approach provides a physical insight that is difficult to get with other standard Maxwell's equation solvers operating in the time or spectral domains.

Using Eqs. (5) and (6), we have computed the optical response of the bowtie antenna under illumination by a normally-incident plane wave polarized along the bowtie axis. A rapid study of the convergence of the QNM expansion as a function of the truncation rank, i.e. the number $M$ of QNMs retained in the expansion, has revealed that no more than 10 QNMs have a significant impact on the extinction and absorption cross-section spectra in the visible. Note that the longitudinal QNMs are not excited since the excitation of bulk plasmons require free electric charges inside resonators and static QNMs have null electric fields inside resonators; the contribution of PML-modes is weak. Figure 2a compares the predictions of the modal method for $M = 12$ (the QNM eigenfrequencies are illustrated in the inset of Fig. 2b) with full-wave frequency data obtained with the classical frequency-domain solver of COMSOL Multiphysics. The agreement is quantitative over the entire spectrum, and can be further improved to reach an absolute accuracy of $10^{-3}$ by including PML-modes in the modal expansion, see Fig. SI. 6 in the SI.

To illustrate the versatility of the QNM expansion, we further consider the bowtie response in the temporal domain. Nanoresonator dynamics are important to explore new spatial and temporal regimes by controlling light-matter interactions with nanometric and femtosecond precisions [39,40]. The spectral QNM expansion of Eq. (5) admits a simple sister form in the temporal domain

$$\mathbf{\Psi}_{\text{sca}}(\mathbf{r},t) = \sum_m \beta_m(t)\widetilde{\mathbf{\Psi}}_m(\mathbf{r}), \qquad (7)$$

with $\beta_m(t) = \int_{-\infty}^{+\infty} \alpha_m(\omega)\exp(-i\omega t)\mathrm{d}\omega$, where $\alpha_m(\omega)$ is weighted by the spectral power density $\mathbf{E}_{\text{inc}}(\omega) = \int_{-\infty}^{+\infty} \mathbf{E}_{\text{inc}}(t)\exp(i\omega t)\,\mathrm{d}t/2\pi$ of the driving pulse $\mathbf{E}_{\text{inc}}(t)$. In our implementation, $\beta_m(t)$ is conveniently computed with a Fast Fourier Transform algorithm. However, to obtain more insight, it is worth performing the calculation analytically by using contour integration. Injecting the expression of $\mathbf{E}_{\text{inc}}(\omega)$ into Eq. (6) to express $\alpha_m(\omega)$ in terms of $\mathbf{E}_{\text{inc}}(t)$, and then using the Cauchy integral formula for the pole at $\omega = \widetilde{\omega}_m$, we obtain

$$\beta_m(t) = <\widetilde{\mathbf{E}}_m^*|i\widetilde{\omega}_m(\varepsilon_m(\widetilde{\omega}_m) - \varepsilon_b)|\mathbf{F}_{\text{inc}}(t)>_{V_{\text{res}}} \exp(-i\widetilde{\omega}_m t) + <\widetilde{\mathbf{E}}_m|\varepsilon_b - \varepsilon_\infty|\mathbf{E}_{\text{inc}}(t)>_{V_{\text{res}}}, \qquad (8)$$

with $\mathbf{F}_{\text{inc}}(t) = \int_{-\infty}^{t} \mathbf{E}_{\text{inc}}(t')\exp(i\widetilde{\omega}_m t')dt'$. Note that Eq. (8) assumes that the background permittivity $\varepsilon_b$ is frequency-independent.

While it may not be the main point of the present work, we would like to stress by the way that the present formalism may also provide a solid theoretical foundation to the temporal coupled mode theory (CMT) of resonators [41], a famous formalism used to model resonator dynamics. Despite its importance and universality, to our knowledge, the temporal CMT still relies nowadays on phenomenological coupling coefficients that are fitted and is restricted to nearly-Hermitian resonators with weak loss and couplings [41-43]. In contrast, the following equation in a CMT form that is easy derived from Eq. (8) by applying the derivative with respect to $t$ on both sides of the equation,

$$d\beta_m(t)/dt = -i\widetilde{\omega}_m\beta_m(t) + \langle\widetilde{\mathbf{E}}_m^*|i\widetilde{\omega}_m(\varepsilon(\widetilde{\omega}_m) - \varepsilon_\infty)|\mathbf{E}_{\text{inc}}(t)\rangle_{V_{\text{res}}} + \langle\widetilde{\mathbf{E}}_m^*|\varepsilon_b - \epsilon_\infty|d\mathbf{E}_{\text{inc}}(t)/dt\rangle_{V_{\text{res}}}, \qquad (9)$$

provides an analytical and rigorous expression of the coupling coefficient.

The upper panel in Fig. 2b shows the temporal evolution of the $x$-component of the electric field at the gap center of the bowtie for a 10-fs plane-wave Gaussian pulse illumination with a central frequency of $0.5\omega_p$ and a bandwidth of $0.058\omega_p$. As evidenced by the comparison with the red curve, they are very accurate, much more than in earlier QNM-expansion works restricted to quasi-static approximations [42] or small truncations numbers [38].

Comparing the computational performance of different methods is difficult. However we can say that the QNM-expansion method is highly effective. The CPU times to compute the spectral or temporal responses in Fig. 2 with an ordinary desktop computer is only 10 min, which are mainly devoted to the QNM computation, the computation of the modal coefficients being very fast comparatively. The present approach thus shows a convincing potential for fast computations in the temporal domain, and represents an interesting alternative to conventional FDTD methods. Additionally, note that any new instance of the driving field parameters, by varying the pulse duration, polarization or incidence angle, requires an entirely new computation with the FDTD methods, whereas owing to analyticity, the additional computation with the present approach just require to perform a 1D Fast Fourier Transform (FFT). Additional details on the mesh size and computational speed as the truncation rank $M$ is varied are provided in the Supplementary Section 5.2.

The temporal QNM-expansion also provides key clues towards understanding nanoresonator dynamics, since the overall response simply results from the superposition of every individual mode response, see Eq. (7). The bottom panel of Fig. 2b shows the time evolution of the excitation coefficients $\beta_m(t)$ of every individual mode obtained by the FFT. To ease the visual analysis, we additionally show the shape of the incident pulse with the shadowed grey curve.

Moreover, from Eq. (8), it is easy to derive an analytic expression of $\beta_m(t)$ for Gaussian pulses, see Eq. (SI.3-38) in the Supplementary Information. Equation (SI.3-38) contains the error function and is not fully transparent for interpretation. Better insight can be achieved by mimicking the Gaussian pulse with a simpler mathematical form $\mathbf{E}_{\text{inc}}(t) = \mathbf{E}_0 \exp(-i\omega_0 t - |t|/\Delta t)$, with $\mathbf{E}_0 = e_0 \exp(i\omega_0 z/c)\,\hat{\mathbf{x}}$ ($e_0$ being a constant) and $\omega_0$ the pulse central-frequency. Then $\beta_m(t)$ becomes

$$\begin{cases} t < 0, \ \beta_m(t) = C_m^- \exp(-i\omega_0 t - |t|/\Delta t), \\ t > 0, \ \beta_m(t) = C_m^+ \exp(-i\omega_0 t - |t|/\Delta t) + (C_m^+ - C_m^-)\exp(-i\widetilde{\omega}_m t), \end{cases} \quad (10)$$

with $C_m^{\pm} = \langle \widetilde{\mathbf{E}}_m^* | \frac{\widetilde{\omega}_m(\varepsilon_m(\widetilde{\omega}_m) - \varepsilon_b)}{\widetilde{\omega}_m - \omega_0 \pm i/\Delta t} | \mathbf{E}_0 \rangle_{V_{\text{res}}}$, $t = 0$ being the arrival time of the pulse peak at the nanoresonator center $z = 0$. For $t > 0$, it is noticeable that the excitation coefficient has two interference terms with time dependence $\omega_0 t$ and $\widetilde{\omega}_m t$. They may result in marked interference features especially if the two terms have comparable amplitudes for $\omega_{\text{beat}} t \sim 1$, where $\omega_{\text{beat}}$ is the beating frequency $\text{Re}(\widetilde{\omega}_m) - \omega_0$. Additionally, for $|\widetilde{\omega}_m - \omega_0| \gg 1/\Delta t$, i.e. for QNM frequencies lying far outside the spectral range of the incident pulse, $|C_m^+ - C_m^-| \ll |C_m^+|, |C_m^-|$, so that $\beta_m(t)$ presents a temporal shape almost identical to that of the incident pulse. These findings well explain the main features of $\beta_m(t)$ observed in Fig. 2b, as detailed below.

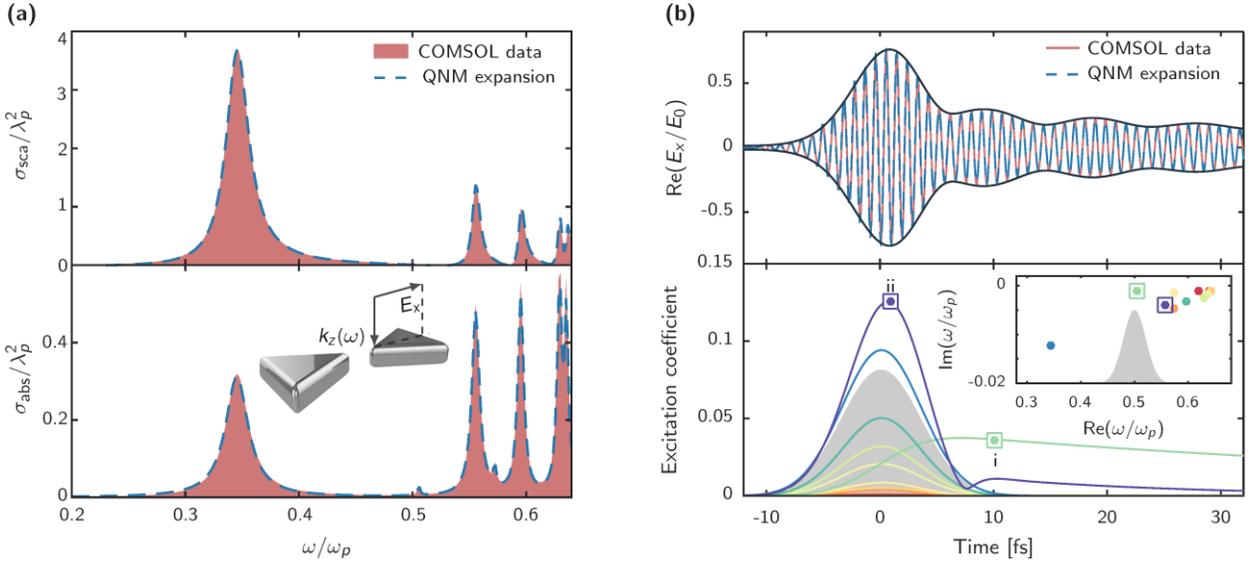

**FIG. 2. Validation of the QNM-expansion formalism in the frequency (a) and time (b) domains.** The results hold for the bowtie illuminated by a plane wave polarized along the $x$-direction with expansions composed of 12 QNMs whose energies and decay rates are shown in the inset in (b). (a) Absorption and scattering cross-section spectra. The maximum absolute error between the QNM-expansion results and "COMSOL data" obtained with frequency-domain simulations is smaller than 0.06. (b) Temporal response for a 10 fs plane-wave Fourier-limited Gaussian pulse, whose temporal and spectral envelopes are shown in the lower panel, with the shadowed grey areas. Top: Temporal evolution of $E_x$ (normalized by the maximum amplitude $E_0$ of the electric field of the incident pulse) at the gap center. The reference COMSOL data are obtained by a Fourier transform of the frequency-domain data. Bottom: Temporal evolution of the QNM excitation coefficients. The two dominant QNMs responsible for the long-tail oscillatory response are highlighted with squares and labeled with "i" and "ii". The inset shows the eigenfrequencies of the 12 QNMs used in the QNM-expansion.

Clearly, the initial bowtie response during the pulse duration $\Delta t$ is due to all the off-resonant QNMs (unlabeled curves) that all exhibit a response that is very similar to the shape of the incident pulse, in agreement with the analysis in the above paragraph. More interesting is the oscillatory long tail response at long times, which is understood as a beating between the contributions of the two QNMs labelled "i" and "ii". The green curve with a long tail results from the energy release of a relatively high-Q ($Q \approx 220$) dark mode with an energy matched with the frequency of the driving pulse. The purple curve of the mode labelled "ii" has a surprising shape, with a kink around 8 fs. Following Eq. (10), the kink can be interpreted as resulting from a destructive interference. This interpretation has been further confirmed by directly examining the companion excitation coefficient $\alpha_{ii}(\omega)$, which exhibits two peaks at the central frequency $0.5\omega_p$ of the incident pulse and the QNM energy $\text{Re}(\widetilde{\omega}_{ii}) = 0.55\omega_p$, as expected from Eq. (10).

The present physical analysis that emphasizes the contribution of every mode in the dynamics contrasts markedly with the black-sensation left by brute-force numerical methods, which are capable of predicting all the fine details of the dynamics but cannot explain their origin. In the spectral domain, the phenomenon related to mode beating in the temporal domain is mode interference that results in a myriad of steep and asymmetric Fano resonance shapes [45]. In our opinion, the QNM expansion of Eq. (5) is the method of choice to analyze and engineer Fano resonances, as it provides a transparent mathematical support that disentangles the essential roles played by the geometry and the driving field parameters in altering the Fano lineshape.

**B. Quenching in nanoresonators**

QNM-expansion formalisms are well known for well predicting the spontaneous emission rate enhancements (Purcell effect) of quantum emitters coupled with the dominant resonance modes of plasmonic antennas [12,42,46,47]. When the emitters approach the metal surfaces down to separation distances smaller than 10 nm, considerable Ohmic heating or quenching is induced at the metal surface just beneath the emitters [48]. Albeit inevitable in plasmonic nanoantennas, quenching has not been previously addressed in the literature on QNM expansions [15], and it appears important to see whether it could be modeled. Incidentally, this will bring us to the important question of the nature of the QNMs responsible for quenching, whose answer will cast doubt on the potential of high-$k$ SPPs. A second motivation for this study is that quenching always occurs at a precise position in a tiny localized volume that strongly depends on the source position, and since QNMs are intrinsic field maps that reflects the symmetry of the geometry and are independent of the source, the modeling of quenching with QNM expansions represents a serious test, which brings us to the question of the limits of the approach.

For clarification, we consider the emission of an on-axis linearly-polarized dipole located at a separation distance $d$ above a silver nanorod in air ($\varepsilon_b = 1$) at a frequency $0.29\omega_p$ matched with the energy of the fundamental dipole resonance of the nanorod. This classical problem has recently received much attention to clarify the connection between the density of electromagnetic states and QNMs [12,14-15]. The numerical predictions obtained with $M = 500$ QNMs are displayed with the dashed-blue curve in Fig. 3a. As $d$ is lowered from 20 nm to 2 nm, the decay rate rapidly increases from 25 to 200, in excellent agreement with fully-vectorial Green-tensor numerical data (red circles) obtained with COMSOL Multiphysics. To illustrate the step forward realized with the present formalism, we also plot the state-of-the-art results (solid-blue curve) obtained in earlier works on QNMs [12,46,47], which were all unable to predict the decay-rate increase for small $d$'s. Consistently, Fig. 3b shows the building-up of the Ohmic loss just beneath the emitter as the truncation rank $M$ increases.

For a single emitter frequency and position, there is no selective advantage to be gained by computing Green function with modal expansions at a single frequency, considering that $\text{M} = 500$ QNMs are needed for recovering accurate results. However, the computation brings important highlights. First the excellent agreement suggests that, at least in the classical dipole-dipole interaction approximations, quenching is indeed a direct consequence of the excitation of high-order plasmon modes. Although widely accepted [49], the role played by high-order plasmons received little evidence for simple 1D geometries [50], or not at all for nanoantennas to our knowledge. Second, it is important to evidence the plasmons involved in the quenching. Their complex frequencies are shown in the right inset of Fig. 3a. Noticeably, quenching arises from a mode accumulation at the surface plasma frequency $\widetilde{\omega}_{SP}$ of high-$k$ SPPs on flat surfaces, defined by $\varepsilon_b = -\varepsilon(\widetilde{\omega}_{SP})$ and identified with a red cross. Intuitively, as $d$ vanishes, the dipole sees a flat interface, and the higher-order plasmons cease to depend on the antenna shape, and resemble those of flat interfaces. Finally, the nature of the modes questions the great virtue attributed to delocalized SPPs on flat surfaces associated to the flat asymptote in the (complex)$\omega$ versus (real)$k$ in all applications related to plasmonic superresolution and confinement [50], since their excitation will be inevitably accompanied with quenching.

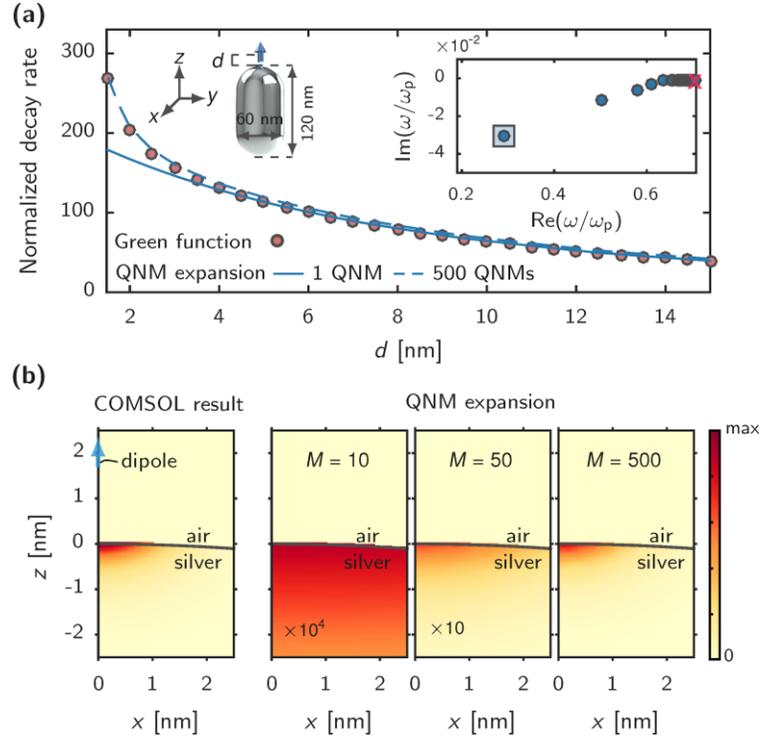

**FIG. 3. Modal analysis of quenching in metallic nanoantennas. (a)** Normalized decay rate of an on-axis-polarized electric-dipole placed at a distance $d$ above a silver nanorod. The dipole emits at the frequency $0.29\omega_p$ matched with the energy of the dominant electric-dipole QNM highlighted by a square in the right inset. The dashed curve computed with $M = 500$ QNMs quantitatively matches the Green-function computational data (red circles) obtained with COMSOL. The solid curve represents the normalized decay rate predicted for $M = 1$ with the dominant electric-dipole QNM, as with all earlier works [12,46,47]. **(b)** Distribution of the absorbed power density $\omega Im(\varepsilon)|\mathbf{E}|^2/2$ computed for a dipole (blue arrow) located at 2 nm above the surface and for several truncation orders $M$. Again, a quantitative agreement is achieved between the distributions computed with the QNM expansion for $M = 500$ (rightmost panel) and COMSOL (leftmost panel), evidencing that the LDOS enhancement due to quenching involves localized-plasmon QNMs, which accumulate at frequencies close to the resonance frequency (highlighted by the red cross in the right inset) of nonretarded slow surface-plasmons on flat interfaces.

### C. Nanoresonators in complex backgrounds: importance of PML-modes

So far, we have considered geometries for which accurate predictions can be achieved with expansions mainly involving QNMs. This is especially possible when the QNM basis is complete, requiring that the Green's tensor is analytic in the complex frequency plane excluding the QNM poles [10,11]. QNM-expansion completeness is guaranteed for 3D nanoresonators in a homogenous background, like in Figs. 1-3, but breaks down for 2D geometries [15,28] or for 3D geometries for which the background is non-uniform, e.g. nanoresonators laying on thin film substrates. Mathematically, the breakdown occurs when the Green's tensor has branch cuts in the complex frequency plane. Since QNMs and PML-modes together constitute a complete basis for the discretized Maxwell's operator in the whole simulation domain (see Section 2.A), when open spaces are mapped onto finite spaces truncated by PMLs, the formal branch cuts of the initial open problem disappear and are replaced by PML-modes, which are expected to carry the extra degrees of freedom.

With the last example, our motivation is to study a complex geometry for which it is necessary to include the PML-modes in the expansion to achieve an accurate reconstruction. The importance of PML-modes for reconstruction has been evidenced so far only for a 2D non-dispersive geometry in air [28], to our knowledge. The geometry consists in a silver nanobullet (a nanorod capped with a hemisphere on one side) laying on a 138-nm-thick semiconductor slab with an infinite spatial extent. We have also tested the case of the same nanobullet on a metal substrate and found that the impact of PML-modes is a little less stringent, see the supplementary Section 5.4.

Figure 4b shows the visible spectrum computed with our QNM-solver in the spectral interval [0.1, 0.7] $\omega_p$. In comparison with Fig. 1a in which PML-modes are simply distributed along a tilted straight line, the spectrum appears more complex. The first salient feature is that it includes several vertical branches that are regularly spaced along the horizontal axis. To intuitively understand their origins, we recall from the discussions in Section 2. A that one class of PML-modes originates from the continuum of background modes that, for Fig. 4, include slab waveguide modes. We believe that the multiple branches are remnant of the PML-transformed waveguide modes of the semiconductor slab that is truncated by PMLs for numerical purpose. Note that similar branch patterns in the complex plane were also observed for 1D gratings due to the existence of different diffraction orders, see Fig. 11 in Ref. [28].

The second salient feature of the spectrum is the entanglement between PML-modes and QNMs, which set close to one another in the complex plane. To discriminate these modes, we exploit the crucial difference between QNMs and PML-modes: QNMs are insensitive to PML-parameter variations, whereas PML-modes are sensitive. Specifically, QNMs and PML-modes are identified by contrasting modes computed with different PMLs. Details are provided in the supplementary Section 3.4.

Figure 4c shows the scattering cross-section $\sigma_{sca}$ computed by neglecting or including PML-modes in the reconstruction. As expected, we observe that by retaining only QNMs in the expansion, it is not possible to accurately predict $\sigma_{sca}$, see the large difference between the blue curve and numerical data obtained with the frequency-solver of COMSOL Multiphysics. Quite the contrary, it is necessary to retain as many as 200 PML-modes in the expansion to achieve a quantitative prediction of the scattering cross-section. The same predictive force is obtained in Fig. 4d for the far field radiation diagram into the free-space modes above and underneath the slab and into the guided modes that are launched into the slab. Details on how the accuracy increases by progressively adding PML-modes in the expansion are provided in the supplementary Section 5.3, see Fig. SI. 7. In this supplementary Section, we additionally study the convergence rate of a similar geometry in which the same nanobullet is placed above a metal substrate (in replacement of the semiconductor slab), see Fig. SI. 8. The intercomparison of the two geometries evidence that the slab case has a slower convergence performance as function of the number of retained eigenmodes in the expansion. We thus consider that the present example, obtained for a 3D dispersive resonator in a non-uniform background and for a complex spectrum that entangle PML-modes and QNMS, constitutes a strong evidence of the generality and soundness of the present method.

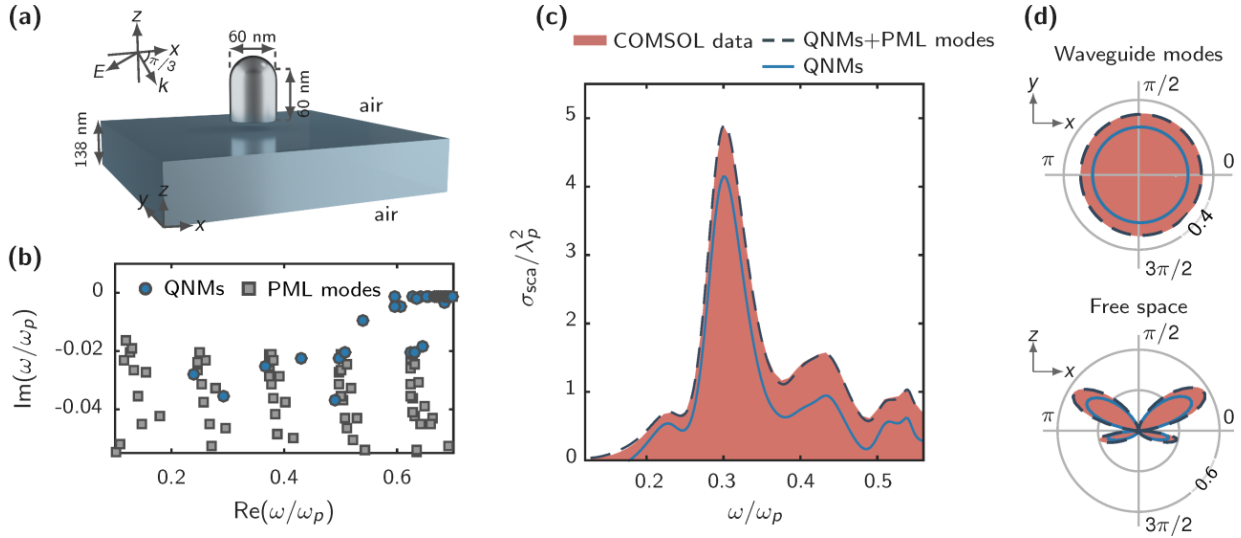

FIG. 4. **Importance of PML-modes. (a)** Schematic representation of a nanobullet on a semiconductor slab slab with an infinite spatial extension in the transverse $x$- and $y$-directions. **(b)** Distribution of the QNM and PML-modes eigenfrequencies in the complex plane. **(c)** Scattering cross-section $\sigma_{sca}$ under illumination by a TM-polarized plane wave incident at oblique angle ($\theta = 60°$ from the $x$-axis). **(d)** Angular distributions ($d\sigma_{sca}/d\Omega\,\lambda_p^2$ in polar diagrams) of the light intensity radiated into the slab guided modes in the transverse $(x,y)$-plane (top) and into free space in the $(x,z)$-plane (bottom) for $\omega = 0.3\omega_p$. We used the near-to-far-field transforms in Ref. [51] to compute the radiation diagrams. In **(c)** and **(d)** the solid-blue and dashed-black curves are computed by retaining 20 QNMs only and 20 QNMs plus 200 PML-modes, respectively, and the shadowed pink curves are the data obtained with the frequency-domain solver of COMSOL Multiphysics. For the reconstruction, the background medium is chosen to be the slab in air, thereby the overlap integrals in Eq. (6) is performed in the nanobullet volume only.

## V. Conclusion

By providing faithful predictions of the dynamics of plasmonic nanoresonators for the general case of complex geometries placed in non-uniform backgrounds, the present QNM-expansion formalism takes an important step toward the deployment of modal theories for analyzing nanoresonators. It features two main building blocks: a FEM-based QNM solver that is robustly applicable for dissipative, dispersive nanoresonators, and a general expression of the modal excitation coefficient for reconstructing scattering fields. The formalism combines the well-known advantages of modal approaches, namely a clarification of the physics as the numerical and physical basis are the same, and an excellent computational performance especially when moderate accuracy is required, when the resonator response is driven by a few dominant modes [15], or when the nanoresonator responses need to be explored for various instances of the driving fields.

Further developments may include refinements of the numerical method, for instance by incorporating dispersive PMLs [1] that allow the computation of QNMs over an extended spectral range. They may also include extensions towards important new applications of nanoresonators at the interface between photonics and other areas of physics, e.g., optomechanical cooling, charge-carrier coupling [52], for which the QNM-expansion formalism is expected to be a smart approach to represent photonic freedoms with dynamically-changing modal coefficients that reveal the complex dynamics of the coupled system.

## Acknowledgements

W.Y. acknowledges a fellowship of the Centre national de la Recherche Scientifique (CNRS). The work was supported by the French National Agency for Research (ANR) under the project "Resonance" (ANR-16-CE24-0013). This study has been carried out with financial support from the French State, managed by the by the French National Agency for Research (ANR ) in the frame of "the Investments for the future " Programme IdEx Bordeaux – LAPHIA (ANR-10-IDEX-03-02). P.L. is pleased to acknowledge the support from the LabEx LAPHIA and from CNRS. We also thank Shanhui Fan for drawing our attention to the potential of the auxiliary-field method. We thank Mondher Besbes, Kevin Vynck, Christophe Sauvan, Boris Gralak, André Nicolet, Marc Duruflé for interesting discussions and computational assistance.

## References

[1] A. Taflove, S.G. Johnson and A. Oskooi, (Artech House, Boston, Mass., 2013).
"Advances in FDTD Computational Electrodynamics: Photonics and Nanotechnology"
[2] J. M. Jin, (Wiley-IEEE Press., 2014).
"The Finite Element Method in Electromagnetics"
[3] A. W. Snyder and J. Love, (Springer, 1983).
"Optical waveguide theory"
[4] D. Marcuse, (Academic Press, 1974).
"Theory of dielectric optical waveguides"
[5] A. J. F. Siegert, Phys. Rev. **56**, 750 (1939).
"On the derivation of the dispersion formula for nuclear reactions"
[6] C. V. Vishveshwara, Nature **227**, 936-938 (1970).
"Scattering of gravitational radiation by a Schwarzschild black-hole"
[7] R. M. More and E. Gerjuoy, Phys. Rev. A **7**, 1288 (1973).
"Properties of Resonance Wave Functions"
[8] B. J. Hoenders, J. Math. Phys. **20**, 329 (1979).
"On the completeness of the natural modes for quantum mechanical potential scattering"
[9] C. E. Baum, Interaction Note 88, December 1971.
See http://www.dtic.mil/cgi-bin/GetTRDoc?Location=U2&doc=GetTRDoc.pdf&AD=ADA066905
"On the singularity expansion method for the solution of electromagnetic interaction problem"
[10] P. T. Leung, S. Y. Liu, and K. Young, Phys. Rev. A **49**, 3057 (1994).
"Completeness and orthogonality of quasinormal modes in leaky optical cavities"
[11] K. M. Lee, P. T. Leung and K. M. Pang, J. Opt. Soc. Am. B **16**, 1409–1417 (1999).
"Dyadic formulation of morphology-dependent resonances. I. Completeness relation"
[12] C. Sauvan, J.-P. Hugonin, I. S. Maksymov and P. Lalanne, Phys. Rev. Lett. **110**, 237401 (2013).
"Theory of the spontaneous optical emission of nanosize photonic and plasmon resonators"
[13] C. Sauvan, J.-P. Hugonin, and P. Lalanne, Proc. SPIE **9546**, Active Photonic Materials VII, 95461C (2015).
"Photonic and plasmonic nanoresonators: a modal approach"
[14] E. A. Muljarov and W. Langbein, Phys. Rev. B **94**, 235438 (2016).
"Exact mode volume and Purcell factor of open optical systems"
[15] P. Lalanne, W. Yan, V. Kevin, C. Sauvan, and J.-P. Hugonin, Laser Photonics Rev. **4**, 1700113 (2018).
"Light interaction with photonic and plasmonic resonances"
[16] E. A. Muljarov, W. Langbein and R. Zimmermann, Europhys. Lett. **92**, 50010 (2010).
"Brillouin-Wigner perturbation theory in open electromagnetic systems"
[17] E. A. Muljarov and W. Langbein, Phys. Rev. B **93**, 075417 (2016).

"Resonant-state expansion of dispersive open optical systems: Creating gold from sand"
[18] J. Yang, H. Giessen, and P. Lalanne, Nano Lett. **15**, 3439 (2015).
"Simple Analytical Expression for the Peak-Frequency Shifts of Plasmonic Resonances for Sensing"
[19] T. Weiss, M. Mesch, M. Schäferling, H. Giessen, W. Langbein, and E.A. Mulfarov, Phys. Rev. Lett. **116**, 237401 (2016).
"From dark to bright: First-order perturbation theory with analytical mode normalization for plasmonic nanoantenna arrays applied to refractive index sensing"
[20] Q. Bai, M. Perrin, C. Sauvan, J.P. Hugonin and P. Lalanne, Opt. Express **21**, 27371 (2013).
"Efficient and intuitive method for the analysis of light scattering by a resonant nanostructure".
[21] J. Zimmerling, L. Wei, P. Urbach and R. Remis, Appl. Phys. A **122**, 158 (2016).
"Efficient computation of the spontaneous decay of arbitrarily shaped 3D nanosized resonators: a Krylov model-order reduction approach"
[22] L. Zschiedrich, F. Binkowski, N. Nikolay, O. Benson, G. Kewes, S. Burger, arXiv:1802.01871 (2018).
"Riesz projection based theory of light-matter interaction in dispersive nanoresonators"
[23] D. A. Powell, Phys. Rev. B **90**, 075108 (2014).
"Resonant dynamics of arbitrarily shaped meta-atoms"
[24] D. A. Powell, Phys. Rev. Applied **7**, 034006 (2017).
"Interference between the modes of an all-dielectric meta-atom"
[25] X. Z. Zheng, V. Volskiy, V. K. Valev, G. A. E. Vandenbosch and V. V. Moshchalkov, IEEE J. Sel. Topics Quantum Electron. **19**, 4600908 (2013).
"Line position and quality factor of plasmonic resonances beyond the quasi-static limit: a full-wave eigenmode analysis route"
[26] R. M. Joseph, S. C. Hagness and A. Taflove, Opt. Lett. **16**, 1412 (1991).
"Direct time integration of Maxwell's equations in linear dispersive media with absorption for scattering and propagation of femtosecond electromagnetic pulses"
[27] A. Raman and S. Fan, Phys. Rev. Lett. **104**, 087401 (2010).
"Photonic band structure of dispersive metamaterials formulated as a Hermitian eigenvalue problem"
[28] B. Vial, A. Nicolet, F. Zolla and M. Commandré, Phys. Rev. A **89**, 023829 (2014).
"Quasimodal expansion of electromagnetic fields in open two-dimensional structures"
[29] N. W. Ashcroft and D. N. Mermin, (Brooks Cole, Belmont, CA, 1976).
"Solid state physics"
[30] F. Olyslager, SIAM J. Appl. Math. **64**, 1408 (2004).
"Discretization of continuous spectra based on perfectly matched layers"
[31] https://www.comsol.com/.
[32] F. Tisseur and K. Meerbergen, SIAM Review **43**, 235-286 (2001).
"The quadratic eigenproblem"
[33] A toolbox package, including a QNM-solver in the form of a COMSOL model sheet and a series of Matlab codes for computing electromagnetic observables with QNMs and PML-modes, is available at the webpage of the authors' group https://www.lp2n.institutoptique.fr/Membres-Services/Responsables-d-equipe/LALANNE-Philippe.
[34] P. B. Johnson and R. W. Christy, Phys. Rev. B **6**, 4370 (1972).
"Optical constants of the noble metals"
[35] See Supplemental Information for detailed derivations and further discussions.
[36] P. T. Leung, S. Y. Liu, and K. Young, Phys. Rev. A 49, 3982 (1994).
"Completeness and time-independent perturbation of the quasinormal modes of an absorptive and leaky cavity"
[37] N. Moiseyev (Cambridge Univ. Press, 2011).
"Non-Hermitian Quantum Mechanics "

[38] R. Faggiani, A. Losquin, J. Yang, E. Mårsell, A. Mikkelsen, P. Lalanne, ACS Photonics **4**, 897 (2017).
"Modal analysis of the ultrafast dynamics of optical nanoresonators"
[39] M. I. Stockman, S. V. Faleev and D. J. Bergman, Phys. Rev. Lett. **88**, 067402 (2002).
"Coherent control of femtosecond energy localization in nanosystems "
[40] L. Piatkowski, N. Accanto and N. F. van Hulst, ACS Photonics **3**, 1401 (2016).
" Ultrafast meets ultrasmall: controlling nanoantennas and molecules "
[41] H. A. Haus and W. Huang, Proc. IEEE **79**, 1505-1518 (1991).
"Coupled-mode theory"
[42] W. Shu, Z. Wang and S. Fan, IEEE J. Quant. Electron. **40**, 1551-1518 (2006).
"Temporal Coupled-Mode Theory and the Presence of Non-Orthogonal Modes in Lossless Multimode Cavities"
[43] S. Fan, W. Suh and J. D. Joannopoulos, J. Opt. Soc. Am. A **20**, 569–572 (2003)
 "Temporal coupled mode theory for Fano resonances in optical resonators"
[44] I. D. Mayergovz, Z. Zhang, and G. Miano, Phys. Rev. Lett. **98**, 147401 (2007).
"Analysis of dynamics of excitation and dephasing of plasmon resonance modes in nanoparticles"
 [45] B. Luk'yanchuk, N. I. Zheludev, S. A. Maier, N. J. Halas, P. Nordlander, H. Giessen, and C. T. Chong, Nat. Mater. **9**, 707 (2010).
"The Fano resonance in plasmonic nanostructures"
[46] P.T. Kristensen, C. van Vlack, and S. Hughes, Opt. Lett. **37**, 1649 (2012).
"Generalized effective mode volume for leaky optical cavities"
[47] R. Ge and S. Hughes, Opt. Lett. **39**, 4235-4238 (2014).
"Design of an efficient single photon source from a metallic nanorod dimer: a quasi-normal mode finite-difference time-domain approach"
[48] K.H. Drexhage, Progress in Optics **12**, 163 (1974).
"Interaction of Light with Monomolecular Dye Layers"
[49]  P. Anger, P. Bharadwaj, and L. Novotny, Phys. Rev. Lett. **96**, 113002 (2006).
" Enhancement and Quenching of Single-Molecule Fluorescence"
[50] A. Archambault, T.V. Teperik, F. Marquier, and J.J. Greffet, Phys. Rev. B **79**, 195414 (2009).
"Surface plasmon Fourier optics"
[51] J. Yang, J.-P. Hugonin and P. Lalanne, ACS Nano **3**, 395 (2016).
"Near-to-Far Field Transformations for Radiative and Guided Waves"
[52] M. Kauranen and A.V. Zayats, Nature Photon. **6**, 737-748 (2012).
"Nonlinear plasmonics"

# SUPPLEMENTARY INFORMATION

# Rigorous Modal Analysis of Plasmonic Nanoresonators

Wei Yan,[1] Rémi Faggiani,[1] and Philippe Lalanne[1]

[1]*Laboratoire Photonique, Numérique et Nanosciences (LP2N), IOGS-Univ. Bordeaux-CNRS, 33400 Talence cedex, France*

**CONTENTS**

The supplementary information (SI) provides a thorough derivation of the analytical formulae presented in the modal formalism, technical details of the implementation of the QNM eigensolver with COMSOL Multiphysics, and an evaluation of the numerical performance of the approach, including the accuracy of the QNM solver and the convergence of the reconstruction as a function of the number of modes retained in the modal expansion.

## 1. DEFINITIONS AND NOTATION

**Material parameters.** We consider non-magnetic materials with the vacuum permeability $\mu_0$. We consider a single-pole Lorentz permittivity,

$$\varepsilon(\omega, \mathbf{r}) = \varepsilon_\infty(\mathbf{r}) - \varepsilon_\infty(\mathbf{r}) \frac{\omega_p^2(\mathbf{r})}{\omega^2 - \omega_0^2(\mathbf{r}) + i\omega\gamma(\mathbf{r})}, \quad (\text{SI.1--1})$$

where the explicit dependence of $\varepsilon_\infty$, $\omega_p$ and $\gamma$ on $\mathbf{r}$ is manifested. The generalization to an $N$-pole Lorentz permittivity, $\varepsilon(\omega, \mathbf{r}) = \varepsilon_\infty(\mathbf{r}) - \varepsilon_\infty(\mathbf{r}) \sum_{i=1}^{N} \omega_{p,i}^2(\mathbf{r})[\omega^2 - \omega_{0,i}^2(\mathbf{r}) + i\omega\gamma_i(\mathbf{r})]^{-1}$, is straightforward.

**Auxiliary fields.** To linearize the source-free Maxwell's equations with respect to the frequency for dispersive materials with a single-pole Lorentz permittivity, we introduce two auxiliary fields [1]:

$$\mathbf{P} = -\varepsilon_\infty \frac{\omega_p^2}{\omega^2 - \omega_0^2 + i\omega\gamma} \mathbf{E}, \quad \mathbf{J} = -i\omega\mathbf{P}, \quad (\text{SI.1--2})$$

where $\mathbf{E}$ denotes the electric field. For an $N$-pole Lorentz permittivity, two auxiliary fields for each Lorentz pole, $\mathbf{P}_i = -\varepsilon_\infty \omega_{p,i}^2 \left(\omega^2 - \omega_{0,i}^2 + i\omega\gamma_i\right)^{-1} \mathbf{E}$ and $\mathbf{J}_i = -i\omega \mathbf{P}_i$ with $i = 1, 2, \cdots, N$, are needed.

**Maxwell's equations with auxiliary fields.** For a single-pole Lorentz permittivity, Maxwell's equations with auxiliary fields read as

$$\hat{\mathbf{H}}(\mathbf{r})\mathbf{\Psi}(\omega, \mathbf{r}) = \omega \mathbf{\Psi}(\omega, \mathbf{r}) + \mathbf{S}(\omega, \mathbf{r}), \quad (\text{SI.1--3})$$

with

$$\hat{\mathbf{H}}(\mathbf{r}) = \begin{bmatrix} 0 & -i\mu_0^{-1}\nabla\times & 0 & 0 \\ i\varepsilon_\infty^{-1}(\mathbf{r})\nabla\times & 0 & 0 & -i\varepsilon_\infty^{-1}(\mathbf{r}) \\ 0 & 0 & 0 & i \\ 0 & i\omega_p^2(\mathbf{r})\varepsilon_\infty(\mathbf{r}) & -i\omega_0^2(\mathbf{r}) & -i\gamma(\mathbf{r}) \end{bmatrix}, \quad \mathbf{\Psi}(\omega, \mathbf{r}) = [\mathbf{H}(\omega, \mathbf{r}), \mathbf{E}(\omega, \mathbf{r}), \mathbf{P}(\omega, \mathbf{r}), \mathbf{J}(\omega, \mathbf{r})]^T,$$

$$(\text{SI.1--4})$$

$\mathbf{S}(\omega, \mathbf{r})$ being the external source term. The augmented electromagnetic vector $\mathbf{\Psi}$ has four components; each component is a vector of a 3-dimensional space, e.g., $\mathbf{E} = \left[E_x, E_y, E_z\right]^T$, where the superscript "T" denotes the transpose operator.

Quasi-normal modes (QNMs) of resonators satisfy the source-free Maxwell's equations:

$$\hat{\mathbf{H}}(\mathbf{r})\widetilde{\boldsymbol{\psi}}_m(\mathbf{r}) = \widetilde{\omega}_m \widetilde{\boldsymbol{\psi}}_m(\mathbf{r}), \quad (\text{SI.1--5})$$

where $\widetilde{\omega}_m$ and $\widetilde{\boldsymbol{\psi}}_m(\mathbf{r}) = \left[\widetilde{\mathbf{H}}_m(\mathbf{r}), \widetilde{\mathbf{E}}_m(\mathbf{r}), \widetilde{\mathbf{P}}_m(\mathbf{r}), \widetilde{\mathbf{J}}_m(\mathbf{r})\right]^T$ represents the eigenfrequency and the eigenvector of the $m^{\text{th}}$ QNM.

**Matrix $\hat{\mathbf{D}}$.** The matrix

$$\hat{\mathbf{D}} = \text{diag}\left[-\mu_0, \varepsilon_\infty, \omega_0^2/(\varepsilon_\infty \omega_p^2), -1/(\varepsilon_\infty \omega_p^2)\right] \quad (\text{SI.1--6})$$

will be used to derive the unconjugated form of the Lorentz reciprocity theorem and the orthogonality relation of QNMs and PML-modes. We emphasize that this matrix is different from the matrix $\hat{\mathbf{A}} = \text{diag}\left[\mu_0, \varepsilon_\infty, \omega_0^2/(\varepsilon_\infty \omega_p^2), 1/(\varepsilon_\infty \omega_p^2)\right]$ used in [1] to define the orthogonality condition for normal modes of a Hermitian, dispersive system, such as lossless metallic photonic crystals. For an $N$-pole Lorentz permittivity, $\hat{\mathbf{D}}$ reads as

$$\hat{\mathbf{D}} = \text{diag}[-\mu_0, \varepsilon_\infty, \omega_{0,1}^2/(\varepsilon_\infty \omega_{p,1}^2), -1/(\varepsilon_\infty \omega_{p,1}^2), \cdots, \omega_{0,N}^2/(\varepsilon_\infty \omega_{p,N}^2), -1/(\varepsilon_\infty \omega_{p,N}^2)]. \quad (\text{SI.1--7})$$

## 2. ELECTROMAGNETIC THEOREMS

In this section, we derive two important classical electromagnetic theorems, the unconjugated form of the Lorentz reciprocity theorem and the Poynting theorem, with auxiliary fields. They will be used to obtain the orthogonality relation of QNMs and PML-modes (Sec. 3.2), and the expressions of the absorption and scattering cross-sections (Sec. 3.6). These two theorems hold for the original continuous operator defined on an unbounded space and for the discretized operator defined on a finite space bounded by PMLs.

### 2.1. Unconjugated form of the Lorentz reciprocity theorem

The derivation of the unconjugated form of the Lorentz reciprocity theorem requires two different solutions of Maxwell's equations, $\hat{\mathbf{H}}\mathbf{\Psi}_1 = \omega_1 \mathbf{\Psi}_1 + \mathbf{S}_1$ and $\hat{\mathbf{H}}\mathbf{\Psi}_2 = \omega_2 \mathbf{\Psi}_2 + \mathbf{S}_2$. Applying the operation $\int_V d^3\mathbf{r} \mathbf{\Psi}_2^{\mathrm{T}} \hat{\mathbf{D}}$ to both sides of $\hat{\mathbf{H}}\mathbf{\Psi}_1 = \omega_1 \mathbf{\Psi}_1 + \mathbf{S}_1$, where $V$ denotes integration volume, we obtain

$$\int_V \mathbf{\Psi}_2^{\mathrm{T}} \hat{\mathbf{D}} \hat{\mathbf{H}} \mathbf{\Psi}_1 d^3\mathbf{r} = \omega_1 \int_V \mathbf{\Psi}_2^{\mathrm{T}} \hat{\mathbf{D}} \mathbf{\Psi}_1 d^3\mathbf{r} + \int_V \mathbf{\Psi}_2^{\mathrm{T}} \hat{\mathbf{D}} \mathbf{S}_1 d^3\mathbf{r}. \tag{SI.2–8}$$

For dielectric materials, the integrals in Eq. (SI.2–8) are simplified by letting $\mathbf{\Psi} = [\mathbf{H}, \mathbf{E}]^{\mathrm{T}}$ and $\hat{\mathbf{D}} = \mathrm{diag}\,[-\mu_0, \varepsilon_\infty]$. With simple, direct algebraic manipulations in the left-hand side of Eq. (SI.2–8), we successively obtain

$$\int_V \mathbf{\Psi}_2^{\mathrm{T}} \hat{\mathbf{D}} \hat{\mathbf{H}} \mathbf{\Psi}_1 d^3\mathbf{r} \stackrel{a}{=} \int_V [\mathbf{H}_2, \mathbf{E}_2, \mathbf{P}_2, \mathbf{J}_2] \begin{bmatrix} 0 & i\nabla\times & 0 & 0 \\ i\nabla\times & 0 & 0 & -i \\ 0 & 0 & 0 & \frac{i\omega_0^2}{\varepsilon_\infty \omega_p^2} \\ 0 & -i & \frac{i\omega_0^2}{\varepsilon_\infty \omega_p^2} & \frac{-i\gamma}{\varepsilon_\infty \omega_p^2} \end{bmatrix} \begin{bmatrix} \mathbf{H}_1 \\ \mathbf{E}_1 \\ \mathbf{P}_1 \\ \mathbf{J}_1 \end{bmatrix} d^3\mathbf{r},$$

$$\stackrel{b}{=} \int_V [\mathbf{H}_1, \mathbf{E}_1, \mathbf{P}_1, \mathbf{J}_1] \begin{bmatrix} 0 & i\nabla\times & 0 & 0 \\ i\nabla\times & 0 & 0 & -i \\ 0 & 0 & 0 & \frac{i\omega_0^2}{\varepsilon_\infty \omega_p^2} \\ 0 & -i & \frac{i\omega_0^2}{\varepsilon_\infty \omega_p^2} & \frac{-i\gamma}{\varepsilon_\infty \omega_p^2} \end{bmatrix} \begin{bmatrix} \mathbf{H}_2 \\ \mathbf{E}_2 \\ \mathbf{P}_2 \\ \mathbf{J}_2 \end{bmatrix} d^3\mathbf{r} + i \int_\Sigma (\mathbf{E}_1 \times \mathbf{H}_2 - \mathbf{E}_2 \times \mathbf{H}_1) \cdot d\mathbf{s},$$

$$= \int_V \mathbf{\Psi}_1^{\mathrm{T}} \hat{\mathbf{D}} \hat{\mathbf{H}} \mathbf{\Psi}_2 + i \int_\Sigma (\mathbf{E}_1 \times \mathbf{H}_2 - \mathbf{E}_2 \times \mathbf{H}_1) \cdot d\mathbf{s},$$

$$\stackrel{c}{=} \omega_2 \int_V \mathbf{\Psi}_1^{\mathrm{T}} \hat{\mathbf{D}} \mathbf{\Psi}_2 + \int_V \mathbf{\Psi}_1^{\mathrm{T}} \hat{\mathbf{D}} \mathbf{S}_2 + i \int_\Sigma (\mathbf{E}_1 \times \mathbf{H}_2 - \mathbf{E}_2 \times \mathbf{H}_1) \cdot d\mathbf{s}, \tag{SI.2–9}$$

where steps a, b and c come from

a. Application of the expression of $\hat{\mathbf{D}}\hat{\mathbf{H}}$.

b. Application of the divergence theorem: for example, $\int_V \mathbf{H}_2 \cdot \nabla \times \mathbf{E}_1 d\mathbf{r} = \int_V \mathbf{E}_1 \cdot \nabla \times \mathbf{H}_2 d\mathbf{r} + \int_\Sigma (\mathbf{E}_1 \times \mathbf{H}_2) \cdot d\mathbf{s}$, $\Sigma$ being the closed surface enclosing $V$.

c. Application of $\hat{\mathbf{H}}\mathbf{\Psi}_2 = \omega_2 \mathbf{\Psi}_2 + \mathbf{S}_2$.

By equating the right-hand side (RHS) of Eq. (SI.2–9) with the RHS of Eq. (SI.2–8), we obtain

$$(\omega_1 - \omega_2) \int_V \mathbf{\Psi}_1^{\mathrm{T}} \hat{\mathbf{D}} \mathbf{\Psi}_2 d^3\mathbf{r} + \int_V \left( \mathbf{\Psi}_2^{\mathrm{T}} \hat{\mathbf{D}} \mathbf{S}_1 - \mathbf{\Psi}_1^{\mathrm{T}} \hat{\mathbf{D}} \mathbf{S}_2 \right) d^3\mathbf{r} = i \int_\Sigma (\mathbf{E}_1 \times \mathbf{H}_2 - \mathbf{E}_2 \times \mathbf{H}_1) \cdot d\mathbf{s}, \tag{SI.2–10}$$

which is the unconjugated form of the Lorentz reciprocity theorem with auxiliary fields.

### 2.2. Poynting theorem

To derive the Poynting theorem, we start from Eq. (SI.1–3), $\hat{\mathbf{H}}\mathbf{\Psi} = \omega \mathbf{\Psi} + \mathbf{S}$. Applying the operation $\int_V d^3\mathbf{r} \mathbf{\Psi}^{\dagger} \hat{\mathbf{A}}$ to both sides of the equation, we obtain

$$\int_V \mathbf{\Psi}^{\dagger} \hat{\mathbf{A}} \hat{\mathbf{H}} \mathbf{\Psi} d^3\mathbf{r} = \omega \int_V \mathbf{\Psi}^{\dagger} \hat{\mathbf{A}} \mathbf{\Psi} d^3\mathbf{r} + \int_V \mathbf{\Psi}^{\dagger} \hat{\mathbf{A}} \mathbf{S} d^3\mathbf{r}, \tag{SI.2–11}$$

where $\mathbf{\Psi}^\dagger$ is the conjugate transpose of $\mathbf{\Psi}$, $\hat{\mathbf{A}} = \text{diag}\left[\mu_0, \varepsilon_\infty, \omega_0^2/\left(\varepsilon_\infty \omega_p^2\right), 1/\left(\varepsilon_\infty \omega_p^2\right)\right]$, and $\omega$ is complex-valued.

Elementary algebraic manipulations on the RHS of Eq. (SI.2–11), similar to those applied to derive Eq. (SI.2–9) from Eq. (SI.2–8), lead to

$$-2\text{Im}(\omega)W_\text{e} = P_\text{abs} + P_\text{rad} - P_\text{inp}, \tag{SI.2–12}$$

where

$$W_\text{e} = \frac{1}{4}\int_V \mu_0|\mathbf{H}|^2 + \varepsilon_\infty|\mathbf{E}|^2 + \frac{1}{\varepsilon_\infty \omega_p^2}\left(|\mathbf{J}|^2 + \omega_0^2|\mathbf{P}|^2\right) d^3\mathbf{r} \quad \text{(energy)}, \tag{SI.2–13a}$$

$$P_\text{inp} = -\frac{1}{2}\int_V \text{Im}\left(\mathbf{\Psi}^\dagger \hat{\mathbf{A}}\mathbf{S}\right) d^3\mathbf{r} \quad \text{(input power)}, \tag{SI.2–13b}$$

$$P_\text{abs} = \frac{1}{2}\int_V \frac{\gamma}{\varepsilon_\infty \omega_p^2}|\mathbf{J}|^2 d^3\mathbf{r} \quad \text{(absorption power)}, \tag{SI.2–13c}$$

$$P_\text{rad} = \frac{1}{2}\int_\Sigma \text{Re}(\mathbf{E} \times \mathbf{H}^*) \cdot d\mathbf{s} \quad \text{(radiation power)}. \tag{SI.2–13d}$$

We note that, in Eq. (SI.2–13a), the sum of the first two terms corresponds to the *electromagnetic energy*, while the last two terms corresponding to the mechanical *kinetic* and *potential* energy of the electron. For dispersive permittivities with multiple Lorentz poles, $W_e$ and $P_\text{abs}$ become

$$W_\text{e} = \frac{1}{4}\int_V \mu_0|\mathbf{H}|^2 + \varepsilon_\infty|\mathbf{E}|^2 + \sum_{i=1}^N \frac{1}{\varepsilon_\infty \omega_{p,i}^2}\left(|\mathbf{J}_i|^2 + \omega_0^2|\mathbf{P}_i|^2\right) d^3\mathbf{r}, \tag{SI.2–14}$$

$$P_\text{abs} = \frac{1}{2}\int_V \sum_{i=1}^N \frac{\gamma_i}{\varepsilon_\infty \omega_{p,i}^2}|\mathbf{J}_i|^2 d^3\mathbf{r}. \tag{SI.2–15}$$

Equation (SI.2–12) is the Poynting theorem, which governs the energy conservation law of electromagnetic systems: the decay rate of the electromagnetic energy, $-2\text{Im}(\omega)W_\text{e}$, equals to the power loss $P_\text{abs} + P_\text{rad}$—arising from material absorption and radiation leakage, respectively—minus the input power $P_\text{inp}$.

We now apply the Poynting theorem Eq. (SI.2–12) to a QNM, with a complex frequency $\widetilde{\omega}$. Since the QNM satisfies the source free Maxwell's equations, $P_\text{inp} = 0$ and Eq. (SI.2–12) becomes $-2\text{Im}(\widetilde{\omega})W_\text{e} = P_\text{abs} + P_\text{rad}$. Remembering the quality factor $Q$ of the QNM can be defined as $Q = -\text{Re}(\widetilde{\omega})/[2\text{Im}(\widetilde{\omega})]$, we obtain

$$Q = -\frac{\text{Re}(\widetilde{\omega})}{2\text{Im}(\widetilde{\omega})} = \text{Re}(\widetilde{\omega})\frac{W_\text{e}}{P_\text{abs} + P_\text{rad}} = 2\pi\frac{\text{Energy stored}}{\text{Energy dissipated by cycle}}, \tag{SI.2–16}$$

which shows that the quality factor is $2\pi$ times the energy stored divided by the energy dissipated per cycle [2]. Interestingly, Eq. (SI.2–16) is generally valid for any closed surface $\Sigma$. The latter may be fully included in the resonator, surround the resonator or be located outside the physical resonator volume. For all cases, the ratio between the energy stored (electromagnetic energy + kinetic energy + potential energy of the electrons) and the power lost by absorption and leakage is a constant, related to the quality factor of the resonator. We however note that the ratio between the absorption and the leakage, $P_\text{abs}/P_\text{rad}$, varies with $\Sigma$. This suggests that the usual decomposition of the resonator-mode $Q$ into two intrinsic contributions, $1/Q = 1/Q_\text{abs} + 1/Q_\text{rad}$ with $Q_\text{abs} = \text{Re}(\widetilde{\omega})W_\text{e}/P_\text{abs}$ and $Q_\text{rad} = \text{Re}(\widetilde{\omega})W_\text{e}/P_\text{rad}$, requires to consider a specific closed surface.

## 3. MODAL FORMALISM: THEORETICAL RESULTS

All the theoretical results of this Section are obtained for QNMs and PML-modes of the discretized operator defined on a finite space bounded by PMLs.

### 3.1. Completeness of QNMs and PML-modes

The completeness of the eigenstates is not always ensured in non-Hermitian systems if the accident degeneracies, the so-called exceptional points, exist, leading to the coalescence of the eigenstates. In the present article, we simply consider the resonator system without the exceptional points. In this case, spectral theory [3; 4] tells us that the completeness relation for eigenstates of a linear non-Hermitian eigenproblem can be constructed by introducing the eigenstates of the complex conjugate transpose of the initial eigenoperator, the so-called adjoint eigenstates. Let us consider the Maxwell's operator $\hat{\mathbf{H}}$ and denote its adjoint operator by $\hat{\mathbf{H}}^\dagger$. Direct evaluation of $\hat{\mathbf{H}}^\dagger \hat{\mathbf{D}}$ and $\hat{\mathbf{D}}\hat{\mathbf{H}}^*$ gives

$$\hat{\mathbf{H}}^\dagger \hat{\mathbf{D}} = \hat{\mathbf{D}}\hat{\mathbf{H}}^*. \tag{SI.3-17}$$

Applying the complex conjugate to the eigenequation Eq. (SI.1–5), we have $\hat{\mathbf{H}}^* \widetilde{\boldsymbol{\psi}}_m^* = \widetilde{\omega}_m^* \widetilde{\boldsymbol{\psi}}_m^*$, which with Eq. (SI.3–17) leads to

$$\hat{\mathbf{H}}^\dagger \hat{\mathbf{D}} \widetilde{\boldsymbol{\psi}}_m^* = \widetilde{\omega}_m^* \widetilde{\boldsymbol{\psi}}_m^*. \tag{SI.3-18}$$

This shows that the vector $\hat{\mathbf{D}}\widetilde{\boldsymbol{\psi}}_m^*$ is an eigenstate of $\hat{\mathbf{H}}^\dagger$ with the eigenvalue $\widetilde{\omega}_m^*$, i.e., the adjoint eigenstate of $\widetilde{\boldsymbol{\psi}}_m$ is $\hat{\mathbf{D}}\widetilde{\boldsymbol{\psi}}_m^*$. $\widetilde{\boldsymbol{\psi}}_m$ and $\hat{\mathbf{D}}\widetilde{\boldsymbol{\psi}}_m^*$ form a biorthogonal basis, as reflected in the orthogonality relation of QNMs and PML-modes, see the next subsection. Accordingly, we have the completeness relation

$$\sum_{m=1}^\infty \widetilde{\boldsymbol{\psi}}_m(\mathbf{r}') \left[\hat{\mathbf{D}}\widetilde{\boldsymbol{\psi}}_m^*(\mathbf{r})\right]^\dagger = \sum_{m=1}^\infty \widetilde{\boldsymbol{\psi}}_m(\mathbf{r}')\widetilde{\boldsymbol{\psi}}_m^{\mathrm{T}}(\mathbf{r})\hat{\mathbf{D}} = \hat{\mathbf{I}}\delta(\mathbf{r}-\mathbf{r}'), \tag{SI.3-19}$$

where $\hat{\mathbf{I}}$ is an identity matrix with the same dimension as $\hat{\mathbf{D}}$.

### 3.2. Orthogonality of QNMs and PML-modes

This section provides the derivation of the orthogonality relation of QNMs and PML-modes in PML mapped space, i.e., Eq. (3) in the main text, by using the unconjugated Lorentz reciprocity theorem presented in Section 2.1.

We consider two eigenmodes of a PML-mapped space, say $\{\widetilde{\omega}_m, \widetilde{\boldsymbol{\psi}}_m\}$ and $\{\widetilde{\omega}_n, \widetilde{\boldsymbol{\psi}}_n\}$, such that $\hat{\mathbf{H}}\widetilde{\boldsymbol{\psi}}_m = \widetilde{\omega}_m\widetilde{\boldsymbol{\psi}}_m$ and $\hat{\mathbf{H}}\widetilde{\boldsymbol{\psi}}_n = \widetilde{\omega}_n\widetilde{\boldsymbol{\psi}}_n$. Plugging these two eigenmodes into Eq. (SI.2–10) and noticing that they satisfy the source-free Maxwell's equations, we obtain

$$(\widetilde{\omega}_m - \widetilde{\omega}_n) \int_V \widetilde{\boldsymbol{\psi}}_m^{\mathrm{T}} \hat{\mathbf{D}} \widetilde{\boldsymbol{\psi}}_n d^3\mathbf{r} = i \int_\Sigma \left(\widetilde{\mathbf{E}}_m \times \widetilde{\mathbf{H}}_n - \widetilde{\mathbf{E}}_n \times \widetilde{\mathbf{H}}_m\right) \cdot d\mathbf{s}. \tag{SI.3-20}$$

We choose the integral domain $V$ to be the whole PML-mapped space, so that $\Sigma$ represents the outer surfaces of PMLs, which are made of perfect electric/magnetic conductors. Perfect electric/magnetic conductors impose zero tangential components of electric/magnetic fields on $\Sigma$. As a result, the surface integral on the RHS of Eq. (SI.3–20) is zero; so, $\widetilde{\boldsymbol{\psi}}_m$ and $\widetilde{\boldsymbol{\psi}}_n$ with $\widetilde{\omega}_m \neq \widetilde{\omega}_n$ satisfy $\int_V \widetilde{\boldsymbol{\psi}}_m^{\mathrm{T}} \hat{\mathbf{D}} \widetilde{\boldsymbol{\psi}}_n d\mathbf{r}^3 = 0$. Therefore, the orthonormal condition of QNMs and PML-modes is defined as

$$\int_V \widetilde{\boldsymbol{\psi}}_m^{\mathrm{T}} \hat{\mathbf{D}} \widetilde{\boldsymbol{\psi}}_n d\mathbf{r}^3 = \delta_{nm}, \tag{SI.3-21}$$

where $\delta_{nm} = 1$ for $n = m$ and 0 otherwise. Equation (SI.3–21) tells us that the eigenstates $\widetilde{\boldsymbol{\psi}}_m$ and their adjoint counterparts $\hat{\mathbf{D}}\widetilde{\boldsymbol{\psi}}_m^*$ form a biorthogonal basis by noticing that $\widetilde{\boldsymbol{\psi}}_m^{\mathrm{T}}\hat{\mathbf{D}} = (\hat{\mathbf{D}}\widetilde{\boldsymbol{\psi}}_m^*)^\dagger$ and accordingly $\int_V (\hat{\mathbf{D}}\widetilde{\boldsymbol{\psi}}_m^*)^\dagger \widetilde{\boldsymbol{\psi}}_n d\mathbf{r}^3 = \delta_{nm}$.

### 3.3. Importance of PML-modes

To understand the role played by PML-modes in the expansion and their impact on the convergence performance of modal expansions, it is relevant to consider the initial unbounded system defined on an open space, see Fig. 1a. For this system, it is established that the QNMs, i.e. the actual natural modes, form a complete set inside the resonator [5–7] for 1D and 3D resonators surrounded by a uniform background. This has been established by showing that (1) the Greens tensor $\mathbf{G}(\mathbf{r},\mathbf{r}',\omega)$ [1] for $\mathbf{r}$ and $\mathbf{r}'$ inside the resonator is analytic everywhere in the complex plane except at the resonance poles, and (2) $\oint_C |\mathbf{G}(\mathbf{r},\mathbf{r}',\omega)/\omega d\omega| \to 0$ as the radius of any circle contour $C$ approaches infinity. For 2D systems or non-uniform backgrounds, the Greens tensor possesses branch cuts, and it is widely accepted that the pole expansion is not complete, as seems to be confirmed by numerical results [4].

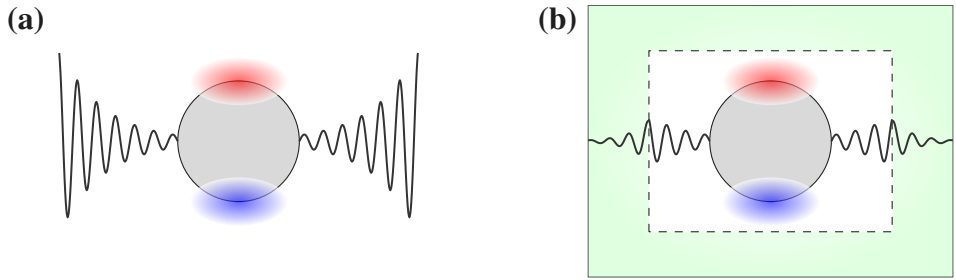

FIG. SI.1  Open space (a) and mapped space (b) truncated with finite-thickness PMLs. (a) The analytic continuation in the complex frequency plane involves poles (the true QNMs) with exponentially-growing fields and potential branch cuts. (b) The discretized operator of the closed mapped space possesses a finite number of eigenstates, including PML-modes that depend on the PML parameters and QNM-like modes (simply called QNMs for terminology simplicity) that do not depend on them and are good approximations of the true QNMs of the open system.

For our mapped operator with finite-thickness PMLs, see Fig. SI.1b, completeness is always guaranteed because the system is closed. As long as the PMLs faithfully satisfy the outgoing-wave conditions over a broad spectral interval around the operating frequencies, the spectrum of the mapped operator is expected to recover the (true) relevant QNMs of the open at least around some operating frequencies. Therefore,

- for a system for which the QNMs do form a complete set, it is expected that accurate predictions can be achieved with expansions retaining only the QNMs of the mapped operator. In practice, as evidenced by the numerical results shown in Figs. 1-3 in the main text, accuracy is often achieved with only QNMs retained in the modal expansion. Note that for such closed systems, completeness is also achieved outside the resonator, if PML-modes are included.

- for a system for which the QNMs do not form a complete set (2D case, complex backgrounds), PML-modes need to be retained in the modal expansion to form a complete set over the whole mapped space. Figure 4 in the main text studies such as situation, for which the presence of a semiconductor slab prevents completeness with QNMs only. Actually, a large amount of PML-modes are required in the expansion for accuracy, as evidenced in Sec. 5.3.

Since the PMLs correctly operate only over a finite spectral range, the true QNMs with eigenfrequencies outside the range are not explicitly recovered as QNM-like eigenstates of the mapped operator, but rather are strongly affected by the PMLs. These modes are classified into PML-modes. In addition, we note that, because of discretization and numerical inaccuracies, the spectrum of the mapped operator contains "numerical modes" with large damping rates. These modes are difficult to differentiate from PML-modes; in general they exhibit field maps with very high spatial frequencies, are weakly excited by the incident field and negligibly impact the reconstruction.

We do not distinguish the numerical modes from PML-modes resulting either from branch cuts or badly-computed QNMs, and call all these modes PML-modes indistinctively.

---

[1] $\mathbf{G}(\mathbf{r},\mathbf{r}',\omega)$ is defined by $\mathbf{\nabla}\times\mathbf{\nabla}\times\mathbf{G}(\mathbf{r},\mathbf{r}',\omega) - \omega^2\mu_0\varepsilon(\mathbf{r},\omega)\mathbf{G}(\mathbf{r},\mathbf{r}',\omega) = \delta(\mathbf{r}-\mathbf{r}')$.

**3.4. Discrimination between QNMs and PML-modes**

Physical interpretation of experimental or computational results is preferentially performed with QNMs. It is thus interesting to consider how discriminating QNMs from PML-modes in the whole set of eigenstates computed with the QNM solver.

For resonators in homogenous backgrounds, the eigenfrequencies of QNMs and PML-modes are located in different regions of the complex plane. In general, we observe a myriad of QNMs close to the real-frequency axis, while PML-modes trace along a tilted straight line, corresponding to the background real-axis continuum rotated by PMLs, e.g., Fig. 1 in the main text.

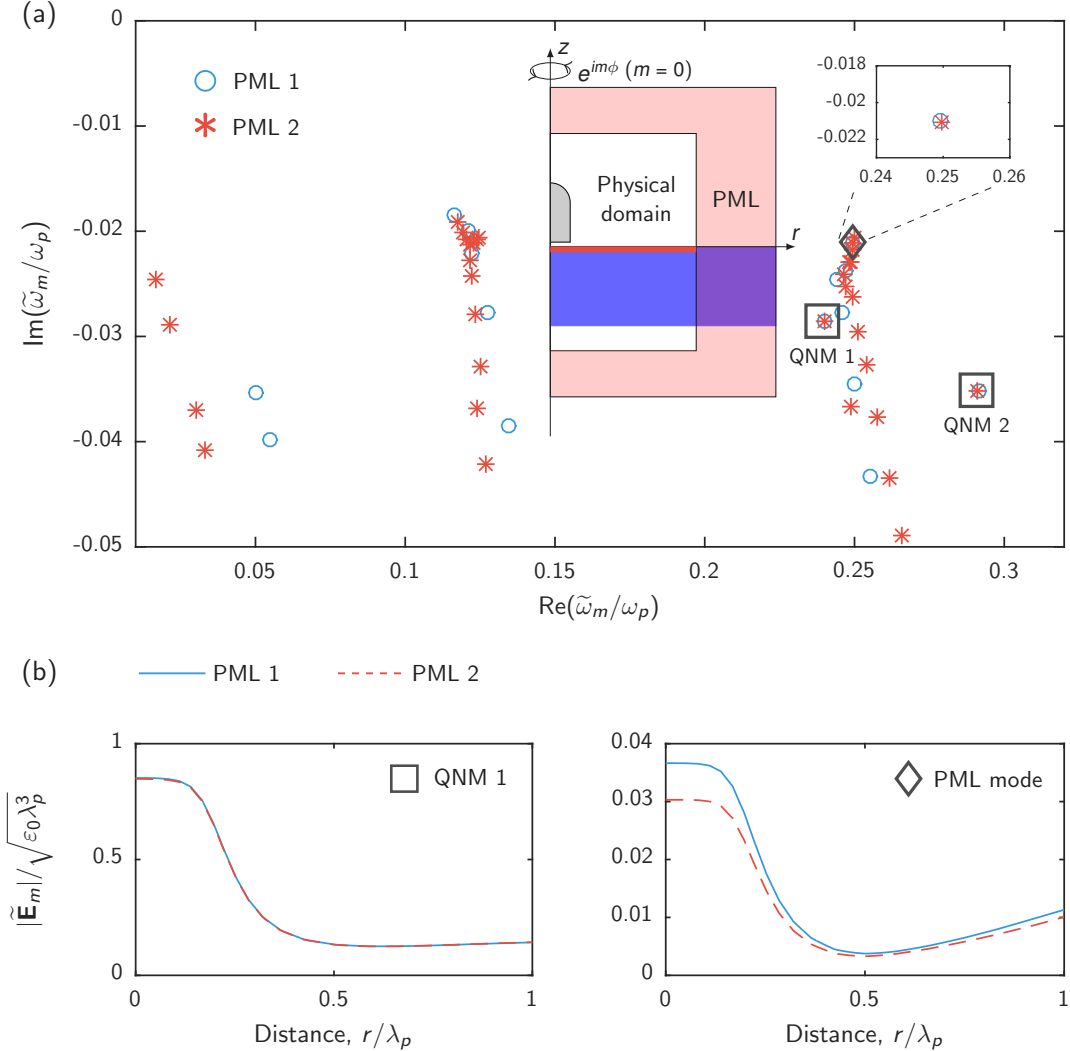

FIG. SI.2 Discrimination between QNMs and PML-modes for a silver nanobullet geometry. The material and geometrical parameters are the same as in Fig. 4 in the main text. (a) Eigenstate energies and decay rates of asymmetrical modes with $m = 0$ computed with the auxiliary-field solver, using two PMLs that have significantly different geometrical sizes and material parameters. Two eigenstates, marked with squares, are identified as QNMs. It may happen by accident that two PML-modes computed with the two PMLs have almost exactly the same eigenfrequencies, see the diamond mark and the associated zoom (upper-right inset). (b) Modulus of the normalized electric field along the thick red line highlighted in the large inset in (a) for QNM1 and for the two PML-modes shown with the diamond mark.

For resonators in inhomogeneous backgrounds, it might be difficult to distinguish between QNMs and PML-modes from their eigenfrequencies solely, see e.g., the nanobullet spectrum of Fig. 4 in the main text. To distinguish them, we may then use the fact that, as discussed in Sec.II A in the main text, QNMs are virtually insensitive to PML-parameter changes, whereas PML-modes are. Figure SI.2(a) illustrates this property for the nanobullet geometry. It shows the spectra computed with the QNM solver using two different PMLs, labelled "PML 1" and "PML 2". For the sake of clarity, only the modes with an azimuthal number $m = 0$ [$\exp(im\phi)$] are shown. The PMLs have nearly identical performance: PML 1 has a thickness twice smaller than that of the PML 2, with an attenuation parameter twice larger. The QNMs are selected by requiring that their eigenfrequencies are the same for both PMLs (we conveniently use a 1% relative variation in our study). As a result, only two QNMs [marked with black squares] are found in the present example in the spectral range of interest. Note that, we might additionally find PML-modes that have almost identical eigenfrequencies, as illustrated with the diamond mark. Nevertheless, the significant change of their mode profiles makes us confident to classify them as PML-modes and not as QNMs, see Fig. SI.2(b).

### 3.5. Modal expansion of the scattered field

In this section, we derive Eq. (6) in the main text, i.e., the closed-form expressions for excitation coefficient of QNMs and PML-modes in the modal expansion of the scattered field. We consider that the resonator is driven by an incident field at the real frequency, which might be radiated by an external electric $\mathbf{J}_0(\mathbf{r})$ or magnetic currents $\mathbf{M}_0(\mathbf{r})$ current. In the scattered field formulation, the permittivity $\varepsilon(\omega, \mathbf{r})$ of the total system [Fig. SI.3(a)] is decomposed as $\varepsilon(\omega, \mathbf{r}) = \varepsilon_b(\omega, \mathbf{r}) + \Delta\varepsilon(\omega, \mathbf{r})$, where $\varepsilon_b$ represents the background permittivity and $\Delta\varepsilon$ is null everywhere except in a small subspace of $\mathbb{R}^3$ that usually defines the resonator volume, denoted by $V_{\text{res}}$ hereafter. Note that $\varepsilon_b$ does not necessarily correspond to a homogeneous medium, as shown in Fig. SI.3(b).

3.5.1. Scattered field formulation

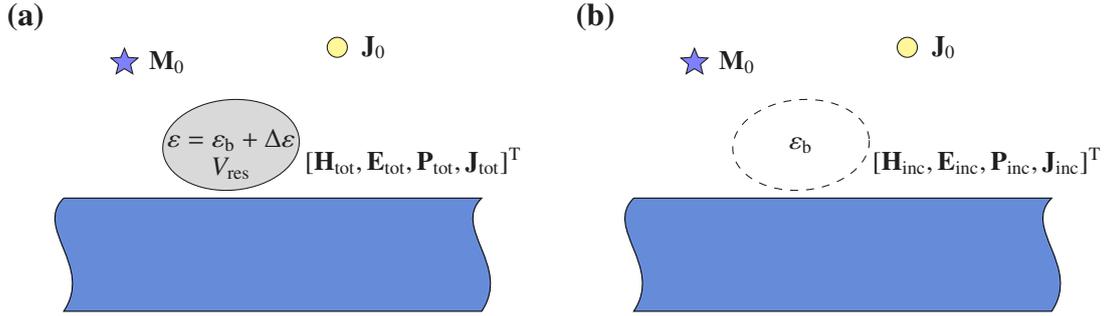

FIG. SI.3 Scattered field formulation. (a) The resonator system, with a permittivity distribution $\varepsilon(\mathbf{r}, \omega)$, driven by external electric currents $\mathbf{J}_0$ and magnetic currents $\mathbf{M}_0$. $\varepsilon(\mathbf{r}, \omega)$ is decomposed as $\varepsilon(\omega, \mathbf{r}) = \varepsilon_b(\omega, \mathbf{r}) + \Delta\varepsilon(\omega, \mathbf{r})$, where $\varepsilon_b$ represents the background permittivity and $\Delta\varepsilon$ is not null in the resonator domain and is null elsewhere. (b) The background medium excluding resonators, with a permittivity distribution $\varepsilon_b(\omega, \mathbf{r})$. The incident field, radiated from the external currents, $\mathbf{J}_0$ and $\mathbf{M}_0$, satisfies Maxwell's equations of the background medium.

Let us briefly recall the traditional scattered field formulation without the auxiliary fields. The incident field [$\mathbf{H}_{\text{inc}}, \mathbf{E}_{\text{inc}}$] satisfies the Maxwell's equations with a background distribution $\varepsilon_b$

$$\nabla \times \mathbf{H}_{\text{inc}} = -i\omega\varepsilon_b \mathbf{E}_{\text{inc}} + \mathbf{J_0}, \quad \nabla \times \mathbf{E}_{\text{inc}} = i\omega\mu_0 \mathbf{H}_{\text{inc}} + \mathbf{M_0}, \qquad (\text{SI.3--22})$$

while the total field [$\mathbf{H}_{\text{tot}}, \mathbf{E}_{\text{tot}}$] satisfies

$$\nabla \times \mathbf{H}_{\text{tot}} = -i\omega\varepsilon \mathbf{E}_{\text{tot}} + \mathbf{J_0}, \quad \nabla \times \mathbf{E}_{\text{tot}} = i\omega\mu_0 \mathbf{H}_{\text{tot}} + \mathbf{M_0}, \qquad (\text{SI.3--23})$$

By difference of Eqs. (SI.3–22) and (SI.3–23), we obtain

$$\nabla \times \mathbf{H}_{\text{sca}} = -i\omega\varepsilon \mathbf{E}_{\text{sca}} - i\omega\Delta\varepsilon \mathbf{E}_{\text{inc}}, \quad \nabla \times \mathbf{E}_{\text{sca}} = i\omega\mu_0 \mathbf{H}_{\text{sca}} \qquad (\text{SI.3--24})$$

The field [$\mathbf{H}_{\text{sca}}, \mathbf{E}_{\text{sca}}$] scattered by the resonant structure at frequency $\omega$ can be seen as the field radiated by a current-source distribution $-i\omega\Delta\varepsilon\mathbf{E}_{\text{inc}}$, a known quantity that is solely depending on the incident driving field.

With auxiliary fields, the augmented electromagnetic vector for the incident field is denoted as $\Psi_{\text{inc}} = [\mathbf{H}_{\text{inc}}, \mathbf{E}_{\text{inc}}, \mathbf{P}_{\text{inc}}, \mathbf{J}_{\text{inc}}]^T$. Moreover, we do not introduce the auxiliary fields for the incident field in the resonator inclusion volume, so that there is $\Psi_{\text{inc}} = [\mathbf{H}_{\text{inc}}, \mathbf{E}_{\text{inc}}, 0, 0]^T$ for $\mathbf{r}$ "belonging to" $V_{\text{res}}$. The augmented electromagnetic vector for the total field, denoted as $\Psi_{\text{tot}} = [\mathbf{H}_{\text{tot}}, \mathbf{E}_{\text{tot}}, \mathbf{P}_{\text{tot}}, \mathbf{J}_{\text{tot}}]^T$, satisfies

$$\hat{\mathcal{H}} \Psi_{\text{tot}} = \begin{bmatrix} 0 & -i\mu_0^{-1}\nabla\times & 0 & 0 \\ i\varepsilon_\infty^{-1}\nabla\times & 0 & 0 & -i\varepsilon_\infty^{-1} \\ 0 & 0 & 0 & i \\ 0 & i\omega_p^2 \varepsilon_\infty & -i\omega_0^2 & -i\gamma \end{bmatrix} \begin{bmatrix} \mathbf{H}_{\text{tot}} \\ \mathbf{E}_{\text{tot}} \\ \mathbf{P}_{\text{tot}} \\ \mathbf{J}_{\text{tot}} \end{bmatrix} = \begin{bmatrix} -i\mu_0^{-1}\mathbf{M}_0 \\ i\varepsilon_\infty^{-1}\mathbf{J}_0 \\ 0 \\ 0 \end{bmatrix}. \tag{SI.3–25}$$

Thus, the augmented electromagnetic vector $\Psi_{\text{sca}} = \Psi_{\text{tot}} - \Psi_{\text{inc}} = [\mathbf{H}_{\text{sca}}, \mathbf{E}_{\text{sca}}, \mathbf{P}_{\text{sca}}, \mathbf{J}_{\text{sca}}]^T$ also satisfies

$$\hat{\mathbf{H}} \Psi_{\text{sca}} = \omega \Psi_{\text{sca}} + \mathbf{S}_{\text{inc}}, \tag{SI.3–26}$$

with

$$\mathbf{S}_{\text{inc}} = [0, \omega \left(\varepsilon_\infty(\mathbf{r}) - \varepsilon_b(\omega, \mathbf{r})\right) / \varepsilon_\infty \mathbf{E}_{\text{inc}}, 0, -i\omega_p^2 \varepsilon_\infty \mathbf{E}_{\text{inc}}]^T \tag{SI.3–27}$$

being the current source induced by the driving field, which is null outside $V_{\text{res}}$.

3.5.2. Modal excitation coefficient

Owing to completeness, $\Psi_{\text{sca}}$ can be expanded everey where in the mapped space by the basis set formed by the QNMs and PML-modes,

$$\Psi_{\text{sca}}(\omega, \mathbf{r}) = \sum_{m=1}^{\infty} \alpha_m(\omega) \widetilde{\psi}_m(\mathbf{r}), \tag{SI.3–28}$$

where $\alpha_m$ is the modal excitation coefficient. To derive a closed-form expression for $\alpha_m$, we apply $\int_V d^3\mathbf{r}\, \Psi_m^T \hat{\mathbf{D}}$ to both sides of Eq. (SI.3–26), plug Eq. (SI.3–28) into Eq. (SI.3–26), then use the orthornormal condition of Eq. (SI.3–21), and finally obtain

$$\begin{aligned} \alpha_m(\omega) &= \frac{\int_V \widetilde{\psi}_m^T \hat{\mathbf{D}} \mathbf{S}_{\text{inc}} d^3\mathbf{r}}{\widetilde{\omega}_m - \omega} \\ &= \frac{\omega}{\widetilde{\omega}_m - \omega} \int_{V_{\text{res}}} [\varepsilon(\widetilde{\omega}_m, \mathbf{r}) - \varepsilon_b(\omega, r)] \widetilde{\mathbf{E}}_m(\mathbf{r}) \cdot \mathbf{E}_{\text{inc}}(\omega, \mathbf{r}) d^3\mathbf{r} + \int_{V_{\text{res}}} [\varepsilon_{\text{res}}(\widetilde{\omega}_m, \mathbf{r}) - \varepsilon_\infty(\mathbf{r})] \widetilde{\mathbf{E}}_m(\mathbf{r}) \cdot \mathbf{E}_{\text{inc}}(\omega, \mathbf{r}) d^3\mathbf{r} \\ &= \frac{\omega}{\widetilde{\omega}_m - \omega} \langle \mathbf{E}_m^*(\mathbf{r})|\varepsilon(\widetilde{\omega}_m, \mathbf{r}) - \varepsilon_b(\omega, \mathbf{r})|\mathbf{E}_{\text{inc}}(\omega, \mathbf{r})\rangle_{V_{\text{res}}} + \langle \mathbf{E}_m^*(\mathbf{r})|\varepsilon(\widetilde{\omega}_m, \mathbf{r}) - \varepsilon_\infty(\mathbf{r})|\mathbf{E}_{\text{inc}}(\omega, \mathbf{r})\rangle_{V_{\text{res}}}, \end{aligned} \tag{SI.3–29}$$

QED. Note that Eq. (SI.3–29) also holds for a dispersive permittivity with multiple Lorentz poles.

### 3.6. Extinction, absorption, scattering, Purcell factor, and temporal response

We summarize in this section the formula used for computing typical electromagnetic observables in nanophotonics including absorption and scattering cross sections, and the Purcell factor.

3.6.1. Scattering and absorption cross sections

We consider a nanoresonator illuminated by a driving field, which in the "resonator-inclusion" volume $V_{\text{res}}$ is $\Psi_{\text{inc}} = [\mathbf{H}_{\text{inc}}, \mathbf{E}_{\text{inc}}, 0, 0]^T$. In the scattered-field formulation, the current source $\mathbf{S}_{\text{inc}}$ generating the scattered field $\Psi_{\text{sca}} = [\mathbf{H}_{\text{sca}}, \mathbf{E}_{\text{sca}}, \mathbf{P}_{\text{sca}}, \mathbf{J}_{\text{sca}}]^T$ is is null every where except in the resonator volume $V_{\text{res}}$, where it is given by $\mathbf{S}_{\text{inc}} = [0, \omega(\varepsilon_\infty - \varepsilon_b)/\varepsilon_\infty \mathbf{E}_{\text{inc}}, 0, -i\omega_p^2 \varepsilon_\infty \mathbf{E}_{\text{inc}}]^T$, see Eq. (SI.3–27). The formulas if the scattering and absorption cross sections are derived by using the Poynting theorem. At a real frequency, it follows from Eq. (SI.2–12) that

$$P_{\text{ext}} = P_{\text{abs}} + P_{\text{rad}}. \tag{SI.3–30}$$

Here, we note that $P_{\text{inp}}$ in Eq. (SI.2–12) is given a different notation, $P_{\text{ext}}$, in agreement with the usual notation for the extinction, i.e. the sum of the scattering and absorption powers. $P_{\text{ext}}$, $P_{\text{abs}}$, and $P_{\text{rad}}$ are given by Eqs. (SI.2–13b)–(SI.2–13d) with the integral volume being $V_{\text{res}}$. The extinction power $P_{\text{ext}}$, i.e., the power supplied by the incident field, is expressed as

$$P_{\text{ext}} = -\frac{1}{2} \int_{V_{\text{res}}} \text{Im} \left[ \omega(\varepsilon_\infty(\mathbf{r}) - \varepsilon_b(\mathbf{r},\omega))\mathbf{E}^*_{\text{sca}}(\mathbf{r}) \cdot \mathbf{E}_{\text{inc}}(\mathbf{r}) - i\mathbf{J}^*_{\text{sca}}(\mathbf{r}) \cdot \mathbf{E}_{\text{inc}}(\mathbf{r}) \right] d\mathbf{r}. \tag{SI.3–31}$$

For a dispersive permittivity with multiple Lorentz poles, the previous equation still holds with $\mathbf{J}_{\text{sca}} = \sum_{i=1}^{N} \mathbf{J}_{\text{sca},i}$ where $\mathbf{J}_{\text{sca},i}$ represents the auxiliary-field current component associated with the $i^{th}$ Lorentz pole.

In practice, we calculate $\mathbf{E}_{\text{sca}}$ and $\mathbf{J}_{\text{sca}}$ with the modal expansion using Eqs. (SI.3–28) and (SI.3–29), and then compute the extinction and absorption cross-sections, $\sigma_{\text{ext}}$ and $\sigma_{\text{abs}}$, using

$$\sigma_{\text{ext}} = \frac{P_{\text{ext}}}{S_0}, \quad \sigma_{\text{abs}} = \frac{P_{\text{abs}}}{S_0}. \tag{SI.3–32}$$

where $S_0$ represents the incident power per unit surface. We then simply calculate the scattering cross-section with $\sigma_{\text{sca}} = \sigma_{\text{ext}} - \sigma_{\text{abs}}$. Note that the computation of the cross-sections just requires to compute a volume integral over the "resonator-inclusion" volume $V_{\text{res}}$.

3.6.2. Purcell factor

Consider an single emitter with an electric dipole moment $\mathbf{p}$ located at the position $\mathbf{r} = \mathbf{r}_0$. Its spontaneous decay rate $\Gamma$ is [8]

$$\Gamma = \frac{2}{\hbar} \text{Im} \left[ \mathbf{p}^* \cdot \mathbf{E}_{\text{tot}}(\mathbf{r}_0) \right], \tag{SI.3–33}$$

where $E_{\text{tot}}$ is the total electric field driven by the dipole. The purcell factor is $\mathcal{F} = \frac{\Gamma}{\Gamma_0}$ with $\Gamma_0$ the decay rate in the background medium. If the background mediums is lossless, isotropic and homogeneous with a relative permittivity $\varepsilon_b$, $\Gamma_0$ has a simple expression $\Gamma_0 = \frac{\omega^3 |\mathbf{p}|^2}{3\pi\hbar c^3 \varepsilon_0} n_b$ with $n_b = \sqrt{\varepsilon_b}$.

In Section 3.5, we derived closed-form expressions for the field scattered by a resonator for an arbitrary illumination. In view of Eq. (SI.3–33), it seems more convenient to represent the total field $\mathbf{\Psi}_{\text{tot}}$ in the QNM basis

$$\mathbf{\Psi}_{\text{tot}} = \sum_m \eta_m \widetilde{\mathbf{\psi}}_m. \tag{SI.3–34}$$

This field $\mathbf{\Psi}_{\text{tot}}$ satisfies Maxwell's equations, Eq. (SI.3–25), with $\mathbf{S}_{\text{inc}} = [0, \omega\mathbf{p}\delta(\mathbf{r} - \mathbf{r}')/\varepsilon_\infty, 0, 0]^T$. Then, $\eta_m$ can be directly obtained by inserting the of $\mathbf{S}_{\text{inc}}$ into the integral in the first line of Eq. (SI.3–29). We then get

$$\eta_m = \frac{\omega \mathbf{p} \cdot \widetilde{\mathbf{E}}_m(\mathbf{r}_0)}{\widetilde{\omega}_m - \omega}, \tag{SI.3–35}$$

and accordingly, the decay rate $\gamma$ reads as

$$\Gamma = \frac{2\omega}{\hbar} \sum_m \text{Im} \left[ \frac{\mathbf{p}^* \cdot \widetilde{\mathbf{E}}_m(\mathbf{r}_0) \otimes \widetilde{\mathbf{E}}_m(\mathbf{r}_0) \cdot \mathbf{p}}{\widetilde{\omega}_m - \omega} \right]. \tag{SI.3–36}$$

Since only the regular part of the Greens tensor has to be considered for calculating the decay rate, uniform convergence is guaranteed by Eq. (SI.3–36) and the usual singularity issues related to convergenge of QNM expansion for the full Greens tensor [9–11] are avoided. We additionally note that, owing to the completeness, Eq. (SI.3–36) holds everywhere in the mapped space, inside the resonator inclusion volume $V_{\text{res}}$ as weel as outside. This contrasts markedly with general results obtained for the QNM expansion in the initial open space, for which completeness is guaranteed only inside the resonator and for simple systems in uniform backgrounds.

### 3.6.3. Radiation diagram

The radiation diagram of resonators for free-space radiation modes and guided modes, as presented in Fig. 4(d) in the main text, is calculated by with the near-to-far field transformation (NFFT) with the near field computed with the modal-expansion method. We refer the reader to Ref. [12–14] for a more complete description of NFFT techniques.

### 3.6.4. Temporal response

In the main text, we have derived the expression of QNM-excitation coefficients in temporal domains, which reads as

$$\beta_m(t) = \left\langle \widetilde{\mathbf{E}}_m^* | i\widetilde{\omega}_m \left(\varepsilon(\widetilde{\omega}_m) - \varepsilon_b\right) | \mathbf{F}_{\text{inc}}(t) \right\rangle_{V_{\text{res}}} \exp(-i\widetilde{\omega}_m t) + \left\langle \widetilde{\mathbf{E}}_m^* | \varepsilon_b - \varepsilon_\infty | \mathbf{E}_{\text{inc}}(t) \right\rangle_{V_{\text{res}}}, \qquad \text{(SI.3–37)}$$

where $\mathbf{F}_{\text{inc}}(t) \equiv \int_{-\infty}^{t} \mathbf{E}_{\text{inc}}(t') dt'$. We now based on Eq. (SI.3–37) derive a more practical expression for $\beta_m(t)$ for Fig. 2(b) in the main text.

Consider that the incident plane wave is a Gaussian pulse, propagating in the $z$ direction with the electric field polarized in the $x$ direction. The incident pulse is thus expressed as $\mathbf{E}_{\text{inc}}(t) = \mathbf{E}_0 \exp(-i\omega_0 t - |t - z/c|^2/\Delta t^2)$ with $\mathbf{E}_0 = e_0 \exp(i\omega_0 z/c)\hat{\mathbf{x}}$ ($e_0$ being a constant). We further assume that the nanoresonator size is much smaller than the pulse spatian extent $c\Delta t$ (for instance, $c\Delta t = 1.5\mu m$ for $\Delta t = 5$ fs ), so that, for a resonator centred at $z = 0$, we could approximate that $\mathbf{E}_{\text{inc}}(t) \approx \mathbf{E}_0 \exp(-i\omega_0 t - t^2/\Delta t^2)$ in Eq. (SI.3–37). Then, a straight froward algebraic calculation from Eq. (SI.3–37) gives us

$$\beta_m(t) = A_m \left[\text{erf}\left(\frac{t + i(\omega_0 - \widetilde{\omega}_m)\Delta t^2/2}{\Delta t}\right) + 1\right] \exp(-i\widetilde{\omega}_m t - (\omega_0 - \widetilde{\omega}_m)^2 \Delta t^2/4) + B_m \exp(-i\omega_0 t - t^2/\Delta t^2), \qquad \text{(SI.3–38)}$$

where

$$A_m = \frac{i\sqrt{\pi}\Delta t}{2} \left\langle \widetilde{\mathbf{E}}_m^* | i\widetilde{\omega}_m \left(\varepsilon(\widetilde{\omega}_m) - \varepsilon_b\right) | \mathbf{E}_0 \right\rangle_{V_{\text{res}}}, \quad B_m = \left\langle \widetilde{\mathbf{E}}_m^* | \varepsilon_b - \varepsilon_\infty | \mathbf{E}_0 \right\rangle_{V_{\text{res}}}, \qquad \text{(SI.3–39)}$$

and erf denotes the error function.

Figure SI.4 compares the predictions of Eq. (SI.3–38) with those obtained with the Fast Fourier Transform algorithm for the two main excitation coefficient of the bow-tie antenna. The agreement is excellent, thereby confirming the validity of Eq. (SI.3–38) for nanoantennas.

## 4. MODAL FORMALISM: IMPLEMENTATION AND NUMERICAL TOOLBOX

### 4.1. Eigenequations and weak formulations

Equation (2) of the main text relies on auxiliary fields to formulate QNMs as a linear eigenequation. QNMs can be computed by descretizing the linear operator $\hat{\mathbf{H}}$ that is defined in PML-mapped systems, and then by solving a standard eigenproblem of a finite-dimensional, linear matrix. So far, the discretization has been mostly implemented with the finite-difference method (FDM) [1; 15; 16]. In this article, for modelling curved boundaries of complex geometries accurately, we employ the finite-element method (FEM) with the commercial COMSOL Multiphysics software and develop an efficient and stable QNM eigensolver, which compute both QNMs and PML-modes.

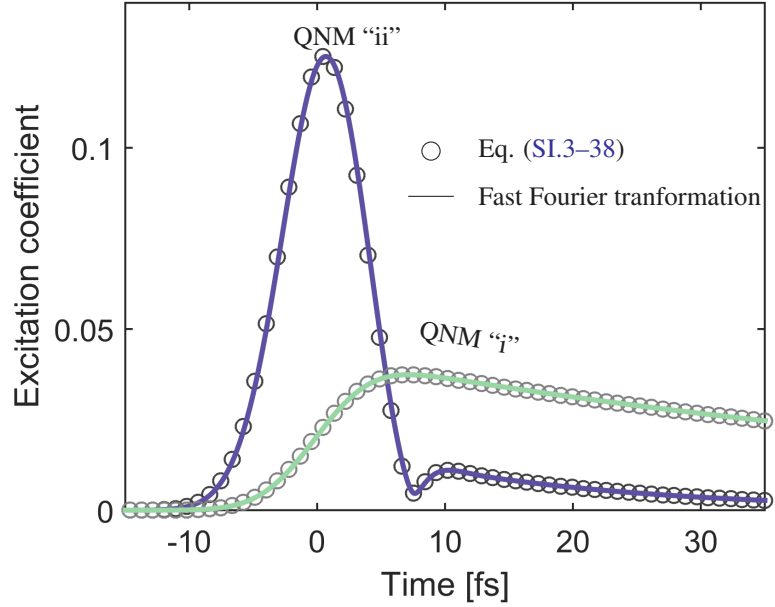

FIG. SI.4 Time-dependence of the excitation coefficients of the QNMs labelled "i" and "ii" for the bow-tie nanoresonator of Fig. 2(b) (main text). Solid line: results computed with a the Fast Fourier Transform algorithm. Circles: results predicted with Eq. (SI.3–38). Note that, since $\varepsilon_b = \varepsilon_\infty$, $B_m = 0$.

Formulating Eq. (2) in the main text into a quadratic form, we first get

$$\underbrace{\begin{bmatrix} \boldsymbol{\nabla}\times\mu^{-1}\boldsymbol{\nabla}\times & 0 \\ \varepsilon_\infty\omega_p^2 & -\omega_0^2 \end{bmatrix}}_{\hat{\mathbf{K}}}\begin{bmatrix}\widetilde{\mathbf{E}}_m \\ \widetilde{\mathbf{P}}_m\end{bmatrix} + \widetilde{\omega}_m\underbrace{\begin{bmatrix} 0 & 0 \\ 0 & i\gamma \end{bmatrix}}_{\hat{\mathbf{C}}}\begin{bmatrix}\widetilde{\mathbf{E}}_m \\ \widetilde{\mathbf{P}}_m\end{bmatrix} + \widetilde{\omega}_m^2\underbrace{\begin{bmatrix} -\varepsilon_\infty & -1 \\ 0 & 1 \end{bmatrix}}_{\hat{\mathbf{M}}}\underbrace{\begin{bmatrix}\widetilde{\mathbf{E}}_m \\ \widetilde{\mathbf{P}}_m\end{bmatrix}}_{\mathbf{u}} = 0, \quad \text{(SI.4–40)}$$

where $\hat{\mathbf{K}}$, $\hat{\mathbf{D}}$, and $\hat{\mathbf{M}}$ are the so-called stiffness, damping, and mass matrices, respectively. Equation (SI.4–40) is suitable for the COMSOL eigensolver that the solves quadratic eigenproblems with a remarkable efficiency through the so-called first companion linearization

$$\begin{bmatrix} \hat{\mathbf{K}} & \hat{\mathbf{C}} \\ 0 & 1 \end{bmatrix}\begin{bmatrix}\mathbf{u}\\ \mathbf{v}\end{bmatrix} = \widetilde{\omega}_m \begin{bmatrix} 0 & -\hat{\mathbf{M}} \\ 1 & 0 \end{bmatrix}\begin{bmatrix}\mathbf{u}\\ \mathbf{v}\end{bmatrix}, \quad \text{with} \quad \mathbf{v} = \widetilde{\omega}_m \mathbf{u}, \quad \text{(SI.4–41)}$$

which is documented, see the survey on the mathematical properties and the numerical solution techniques for solving quadratic eigenvalue problem [17].

The weak form of Eq. (SI.4–40) is

$$\int_V \boldsymbol{\nabla}\times\mathbf{F}(\mathbf{r})\cdot\mu^{-1}\boldsymbol{\nabla}\times\widetilde{\mathbf{E}}_m(\mathbf{r}) - \widetilde{\omega}_m^2 \varepsilon_\infty \mathbf{F}(\mathbf{r})\cdot\widetilde{\mathbf{E}}_m(\mathbf{r}) - \widetilde{\omega}_m^2(\mathbf{r})\mathbf{F}(\mathbf{r})\cdot\widetilde{\mathbf{P}}_m(\mathbf{r})\, d^3\mathbf{r} = 0, \quad \text{(SI.4–42a)}$$

$$\int_V \varepsilon_\infty\omega_p^2 \mathbf{F}(\mathbf{r})\cdot\widetilde{\mathbf{E}}_m(\mathbf{r}) - \omega_0^2 \mathbf{F}(\mathbf{r})\cdot\widetilde{\mathbf{P}}_m(\mathbf{r}) + i\widetilde{\omega}_m\gamma \mathbf{F}(\mathbf{r})\cdot\widetilde{\mathbf{P}}_m(\mathbf{r}) + \widetilde{\omega}_m^2 \mathbf{F}(\mathbf{r})\cdot\widetilde{\mathbf{P}}_m(\mathbf{r})\, d^3\mathbf{r} = 0, \quad \text{(SI.4–42b)}$$

where $\mathbf{F}(\mathbf{r})$ is an arbitrary smooth function and is usually called test function. In the toolbox package, Eqs. (SI.4–42a) and (SI.4–42b) are directly implemented in the weak-form environment of the COMSOL Multiphysics. They are then solved by the build-in eigensolver.

For dielectric resonators made of non dispersive materials, the auxiliary field $\mathbf{P}$ is irrelevant, so is Eq. (SI.4–42b); Equation (SI.4–42a) simplifies to

$$\int_V \boldsymbol{\nabla}\times\mathbf{F}(\mathbf{r})\cdot\mu^{-1}\boldsymbol{\nabla}\times\widetilde{\mathbf{E}}_m(\mathbf{r}) - \widetilde{\omega}_m^2 \varepsilon_\infty \mathbf{F}(\mathbf{r})\cdot\widetilde{\mathbf{E}}_m(\mathbf{r}) d^3\mathbf{r} = 0. \quad \text{(SI.4–43)}$$

For dispersive permittivities with multiple Lorentz poles, Eqs. (SI.4–42a) and (SI.4–42b) become

$$\int_V \boldsymbol{\nabla} \times \mathbf{F}(\mathbf{r}) \cdot \mu^{-1} \boldsymbol{\nabla} \times \widetilde{\mathbf{E}}_m(\mathbf{r}) - \widetilde{\omega}_m^2 \varepsilon_\infty \mathbf{F}(\mathbf{r}) \cdot \widetilde{\mathbf{E}}_m(\mathbf{r}) - \widetilde{\omega}_m^2 \mathbf{F}(\mathbf{r}) \cdot \sum_i^N \widetilde{\mathbf{P}}_{m,i}(\mathbf{r}) \, d^3\mathbf{r} = 0, \quad \text{(SI.4–44a)}$$

$$\int_V \varepsilon_\infty \omega_p^2 \mathbf{F}(\mathbf{r}) \cdot \widetilde{\mathbf{E}}_m(\mathbf{r}) - \omega_0^2 \mathbf{F}(\mathbf{r}) \cdot \widetilde{\mathbf{P}}_{m,i}(\mathbf{r}) + i\widetilde{\omega}_m \gamma \mathbf{F}(\mathbf{r}) \cdot \widetilde{\mathbf{P}}_{m,i}(\mathbf{r}) + \widetilde{\omega}_m^2 \mathbf{F}(\mathbf{r}) \cdot \widetilde{\mathbf{P}}_{m,i}(\mathbf{r}) \, d^3\mathbf{r} = 0. \quad \text{(SI.4–44b)}$$

### 4.2. Eigensolver implementations and toolbox package

We have developed a toolbox package that gathers the most significant developments achieved in the present work. The toolbox is available at *www.lp2n.institutoptique.fr/Membres-Services/Responsables-d-equipe/LALANNE-Philippe*. The toolbox package consists of

1. A QNM eigensolver that takes the form of a COMSOL model sheet. Simply, by drawing a new geometry and defining new materials, the user may compute all the QNMs and PML-modes of his/her own resonator geometries.

2. A series of Matlab codes that use the eigenstates computed with the QNM eigensolver to reconstruct electromagnetic observables, such as field distributions, cross-section spectra, radiation-diagram, temporal-domain responses.

Some important features of the QNM eigensolver are

- The computational domain is a finite space bounded by PMLs. We have Cartesian-, cylindrical- and spherical-coordinate stretched PMLs with anisotropic material parameters, and choose the complex coordinate transformation functions that determine the material parameters of PMLs to be frequency independent. This choice is motivated by simplicity reasons, and there is little doubt that better performance can be achieved with dispersive PMLs [18], for which outgoing waves might be efficiently damped for a much wider frequency range.

- Auxiliary fields are only needed in domains where material parameters are dispersive. If PMLs are used to absorb outgoing waves from dispersive background media, then the material parameters of PMLs also become dispersive, so that auxiliary fields are required in the PML domain.

- The QNM eigensolver is developed in the COMSOL Weak Form PDE module. The weak formulations of Eqs. (SI.4–42a) and (SI.4–42b) are directly input in COMSOL Multiphysics software. The basis function (the so-called shape function in COMSOL Multiphysics) that interpolates solutions among discretized mesh nodes, is chosen to be "curl type"; this type handles the discontinuities of the normal component of electromagnetic fields across boundaries between two different media.

- The build-in eigensolver of COMSOL Multiphysics is used. We use the direct preconditioner for the matrix preconditioning in all the numerical examples shown in this article.

### 4.3. Accuracy of the QNM eigensolver

To evidence the precision reached by the present QNM eigensolver, we consider a silver bowtie antenna studied in the main text with a Drude permittivity, see Table I. We first compute the dominant QNMs at visible frequencies with the QNM eigensolver. The real and imaginary parts of the eigenfrequencies of the four QNMs are given in the second and third columns of the Table I.

For testing the accuracy of the computed data, we further compare these data with those obtained with a freeware dedicated to the computation of the eigenfrequencies of dispersive nanoresonators [19]. The freeware implements a method that relies on an iterative pole search approach to fit representative electromagnetic quantities — e.g., the scattered electric field — to a Padé approximated pole-like response function. The iterative procedure is carried out utilizing the COMSOL Multiphysics solver driven by a MATLAB code. To avoid numerical dispersion, we use exactly the same mesh and the same PMLs as those implemented for the QNM eigensolver. The data computed with the pole search are shown in the fourth and fifth columns of the table. An impressive 5-6 digit agreement for both the real and imaginary parts of the four eigenfrequencies is achieved, as outlined by bold figures in the numbers of the table.

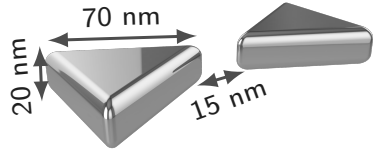

| QNM | Present QNM-eigensolver | | Iterative pole approach | |
|---|---|---|---|---|
| | Re($\widetilde{\omega}$)/$\omega_p$ | Im($\widetilde{\omega}$)/$\omega_p$ | Re($\widetilde{\omega}$)/$\omega_p$ | Im($\widetilde{\omega}$)/$\omega_p$ |
| 1 | **0.34533**001 | **−0.01218966** | **0.34532**940 | **−0.0121895**7 |
| 2 | **0.50558**795 | **−0.00114938** | **0.50558**509 | **−0.00114938** |
| 3 | **0.55559**726 | **−0.00412**330 | **0.55559**427 | **−0.00412**287 |
| 4 | **0.57231**720 | **−0.00138**475 | **0.57231**028 | **−0.00138**585 |

TABLE I  Test of the accuracy of the QNM-solver for the bowtie nanoantenna. The bold figures outline the common digits computed with the two approaches. The values of the eigenstate energies and decay rates are normalized by $\omega_p$.

It is worth emphasizing that, even if the two approaches share many common features (they are computed with the same electromagnetic software, mesh and PMLs), they use totally different algorithms. The pole-search method relies on the standard COMSOL Multiphysics solver; QNMs are computed one by one, and each computation is iterative. In contrast, the QNM eigensolver is implemented in the weak-form environment of COMSOL Multiphysics and computes all the eigenstates in parallel in a much faster way. Thus the 5-6 digit accuracy can be considered as a serious test for the present QNM eigensolver.

## 5. NUMERICAL TESTS

In this section, we study the exactness, the convergence performance and the computational speed of the modal-expansion method.

The exactness is related to the capability of the method to solve Maxwells equations in a rigorous way. To test this capability, we consider a simple geometry, a metallic sphere in air, for which an exact solution is available from Mie's scattering theory. In Sec. 5.1, we evidence that the modal method approaches the exact value of the extinction cross section with a high accuracy, limited only by finite element discretization.

The convergence performance concerns how the accuracy increases as the number $M$ (the truncation rank) of the modes retained in the computation increases. The convergence performance of the method is first studied for the metallic sphere in Sec. 5.1, and then for more complicated geometries, the bowtie antenna in air (Fig. 1 in the main text) in Sec. 5.2, the nanobullet on a semiconductor slab (Fig. 4 in the main text) in Sec. 5.3, and finally the nanobullet on a semi-infinite metallic substrate in Sec. 5.4.

At last, the computational speed of the approach is discussed in Sec. 5.5.

To study the convergence rate, we need to sort the eigenstates to include them progressively one after the other in the expansion, to further look at the increase of the accuracy as the number of modes $M$ retained increases. The basic idea for sorting is to start by first considering the dominant QNMs, which are in general easily recognized in the spectrum computed with the QNM solver. This is indeed what we do, and in general this is also what could be done by experimentalists when interpreting their measurements. However to study the convergence performance at high accuracies (relative errors below $10^{-2} - 10^{-3}$), PML-modes have to be incorporated into the expansion. It is not easy to sort those modes with a straightforward criteria. Thus we consider an objective criteria for which the eigenstates are sorted by increasing order of their impact on the reconstruction. To be more specific, imagine that we are interested in the extinction cross section $\sigma_{ext}$ of a nanoresonator (like in the following tests), we first compute the contribution of each individual eigenstate $m$ to $\sigma_{ext}$ (this can be done analytically using Eq. SI.332), denoted by $\sigma_{ext,m}$, then spectrally average the modulus of $\sigma_{ext,m}$ over some spectral interval $[\omega_1; \omega_2]$ that, in general, corresponds to the spectral domain of interest in the study. Hereafter we denote this spectral average by $<|\sigma_{ext,m}|>_{avg.}$.

**5.1. Exactness: an analytic-solvable example, sphere in air.**

We consider an analytic-solvable example, a metallic sphere in air, which is not studied in the main text. The sphere has a Drude permittivity $\varepsilon = 3 - \omega_p^2/(\omega^2 + i\omega\gamma)$ with $\gamma = 0.0023\omega_p$, where $\lambda_p \equiv 2\pi c/\omega_p = 138$ nm, and has a radius $R = 30$ nm.

Figure SI.5(a) shows the computed eigenfrequency spectrum. For the sake of clarity, we only show axially-symmetric modes with azimuthal number $m = 1$. The spectrum contains pairs of modes with nearly opposite eigenfrequencies [they would be exactly opposite if the loss in the Drude model is neglected, i.e. if $\varepsilon(\omega) = \varepsilon(-\omega)$]. For this simple geometry, QNMs can be exactly computed by finding the poles of Mie's scattering coefficients, and we use this knowledge to distinguish QNMs from PML-modes.

All the modes are sorted (and colored in the figure) by increasing order of their impact on the reconstruction using the spectral range $[\omega_1; \omega_2] = [0.2\omega_p; 0.6\omega_p]$. We note that the the fundamental electric dipole QNM with the resonance frequency around $0.4\omega_p$ has the largest impact, $\approx 8$ times larger than the mode with the second largest weight. We also note that, except for a minority of them, the PML-modes have extremely small weights implying that they negligibly impact the reconstruction.

To study the exactness of the modal method, we first compute the extinction cross section with Mie's theory. The results are shown with circles in Fig. SI.5(b) and are compared with the results obtained with the modal-expansion method for two meshes. The coarse one has a mesh size $h$ ($h \approx 0.2R$ in sphere and $h \approx 3.2R$ in air and PMLs) and the second mesh is finer with a mesh size $h/5$. For the thinner mesh, memory requirement exceeds the capacities of our desktop computer and we use 2D simulations, taking advantage of the axial symmetry.

The extinction cross-section spectra reconstructed with the modal method for $M = 2$, 12 and 20 modes with the coarse mesh are shown with solid curves in Figure SI.5(b). We see that the extinction cross-section spectrum reconstructed with only the two dominant QNMs provides very accurate predictions. By further increasing $M$ and considering higher-order QNMs and PML-modes, the numerical accuracy is steadily improved.

Figure SI.5(c) shows $<|\sigma_{\text{ext}}^{\text{modal}}(\omega) - \sigma_{\text{ext}}^{\text{Mie}}(\omega)|>_{\text{avg}}$, the numerical error of the spectrally-averaged extinction cross-section, where $\sigma_{\text{ext}}^{\text{modal}}$ and $\sigma_{\text{ext}}^{\text{Mie}}$ denote the extinction cross sections obtained with the modal method and Mie's scattering theory, respectively. The spectral averaging is performed from $0.2\omega_p$ to $0.6\omega_p$. To disentangle the errors introduced by the FEM discretization from those due to the truncation, we have also computed the extinction cross-section (dashed lines) with the standard frequency-domain electromagnetic solver of COMSOL Multiphysics using the same meshes. Three important observations can be made: (1) as $M$ increases, the numerical errors progressively reduce to reach a plateau; (2) the value of the plateau almost coincides with the errors achieved with the frequency-domain electromagnetic solver of COMSOL Multiphysics; (3) the absolute accuracy strongly increases from $10^{-3}$ to $10^{-5}$ as the mesh resolution is increased, implying that the numerical inaccuracy of the modal method can be significantly improved by using finer numerical discretization. This evidences that the present modal method can achieve a high level of precision limited by the discretization, just like other classical FEM methods, for simple geometries.

For the sake of completeness, we note the presence of two accumulation points in the spectrum, near the pole frequency of the Drude permittivity ($\omega = -i\gamma$) and near the resonance frequency of slow surface plasmons on flat interfaces [$\varepsilon(\omega) + \varepsilon_b = 0$]. The modes computed with the QNM solver close to these points are included in the convergence rate shown in Fig. SI.5(c).

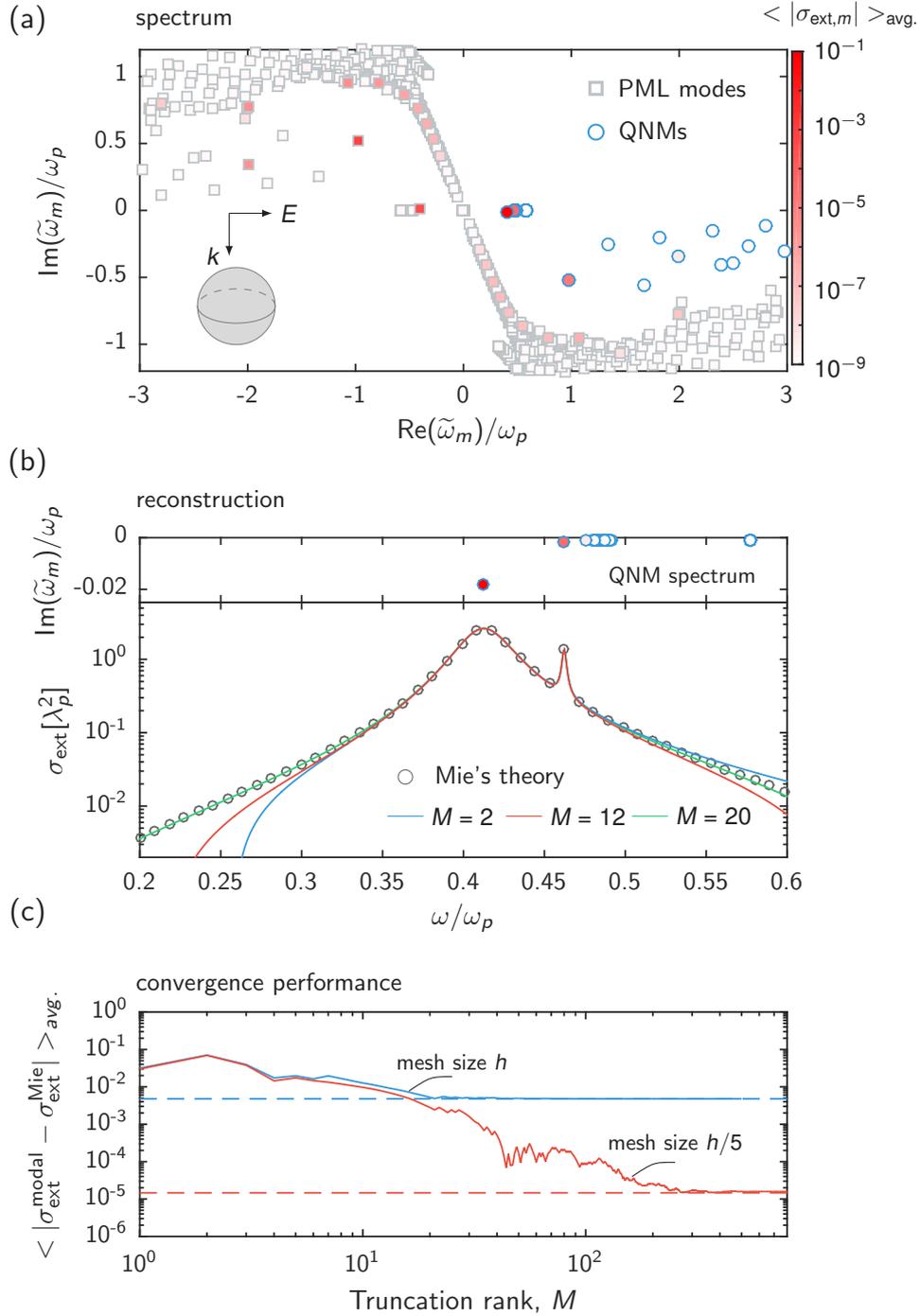

FIG. SI.5 Numerical performance of the modal-expansion method for a metallic sphere in air. The sphere has a Drude permittivity $\epsilon = 3 - \omega_p^2/(\omega^2 + i\omega\gamma)$ with $\gamma = 0.02\omega_p$ with $\lambda_p = 2\pi c/\omega_p = 138$ nm and a radius $R = 30$ nm. (a) Eigenfrequencies of QNMs (circles) and PML-modes (squares) computed with the QNM solver. The color of the markers visualizes $<|\sigma_{\text{ext},m}|>_{\text{avg.}}$, i.e., the contribution of each mode to the spectrally-averaged extinction cross-section. (b) Extinction cross-section spectrum computed with the modal method for $M = 2, 12, 20$ modes and Mie's scattering theory (gray circles). For $M = 2$, we includes the two dominant QNMs, whose resonance frequencies (see top panel) dictate the position and the width of two main resonance peaks. (c) Convergence rate of the modal method. The modes are sorted by increasing order of their impact on the reconstruction. The dash horizontal lines represent the numerical errors of the extinction cross section computed with the standard frequency-domain electromagnetic solver of COMSOL Multiphysics using the same meshes and the same PMLs.

**5.2. Convergence performance: bowtie nanoresonator.**

Figure SI.6 shows the convergence and accuracy of the modal-expansion method for the bowtie nanoresonator studied in Figs. 1 and 2 in the main text. We adopt the same presentation [(a): spectrum, (b): reconstruction, and (c): convergence performance] as that used for the metallic sphere. However, since no analytical result is available for the bowtie, we compare the predictions of the modal method with those obtained with the standard frequency-domain electromagnetic solver (classical solver) of COMSOL Multiphysics. Similar observations as in Fig. SI.5 can be made. We note that, in Fig. SI.6 (c), $< |\sigma_{\text{ext}}^{\text{modal}} - \sigma_{\text{ext}}^{\text{COM}}| >_{\text{avg.}}$, the spectral-average difference of the extinction cross section obtained with the modal method and with the COMSOL frequency-domain solver, approaches a highly-accurate value $\sim 10^{-3} \lambda_p^2$ for $M \approx 100$. This value is one order of magnitude smaller than that shown in Fig. 2 in the main text, where only 12 QNMs are retained in the expansion. Moreover, we observe (not demonstrated here) that the accuracy can be further increased by including more modes accumulated near the frequency for which $\varepsilon(\omega) + \varepsilon_b = 0$, i.e., slow surface plasmon modes.

**5.3. Convergence performance: nanobullet resonator on a semiconductor slab.**

We now consider the geometry studied in Fig. 4 in the main text, a nanobullet resonator laying on a semiconductor slab. The convergence and accuracy of the modal method for this geometry are shown in Fig. SI.7. The PML spectrum [Fig. SI.7 (a)] is more complicated than for the previous geometries. Zooming in the spectrum [Fig.4 (a) in the main text], we see many PML-modes organized along vertical branches. They are PML-transformed waveguide modes of the semiconductor slab, see details in Sec.IV.C in the main text.

For the two previous geometries, it was possible to obtain very accurate results for the cross-section spectra by considering only a few QNMs in the expansions, a key advantage of the present modal approach over other numerical methods. For the nanobullet geometry, because the PML-mode spectrum is complicated, the accuracy achieved by retaining only QNMs [blue curve in Fig. SI.7 (b)] is rather weak; only the overall shape of the spectra in Fig. SI.7 (b) is recovered with a few QNM, but the predictions are not qualitative. Good accuracy (below 1% relative error) is achieved for $M > 100$, see Fig. SI.7 (c).

**5.4. Convergence performance: nanobullet resonator on a semi-infinite metallic substrate**

The geometry studied in this subsection is exactly the same as the previous one, except that the nanobullet is not located on a semiconductor slab, but on a metallic semi-infinite substrate. The latter has the same permittivity as the nanobullet. Figure SI.8 shows the convergence and the accuracy of the modal method. Compared to Fig. SI.7(a), the PML spectrum appears simpler, since many branches of PML-transformed slab waveguide modes no longer exist. Instead, the surface plasmon modes, which are supported by the flat metallic interface, give rise to new branches of PML-modes, which are attached to the real-frequency axis at the resonance frequency of non-retarded slow surface plasmons on flat interfaces, i.e., around $\pm \omega_p / \sqrt{2}$. Moreover, we observe two other branches emerging around $\pm \omega_p$, which are the PML-transformed continuum radiation modes of the metal.

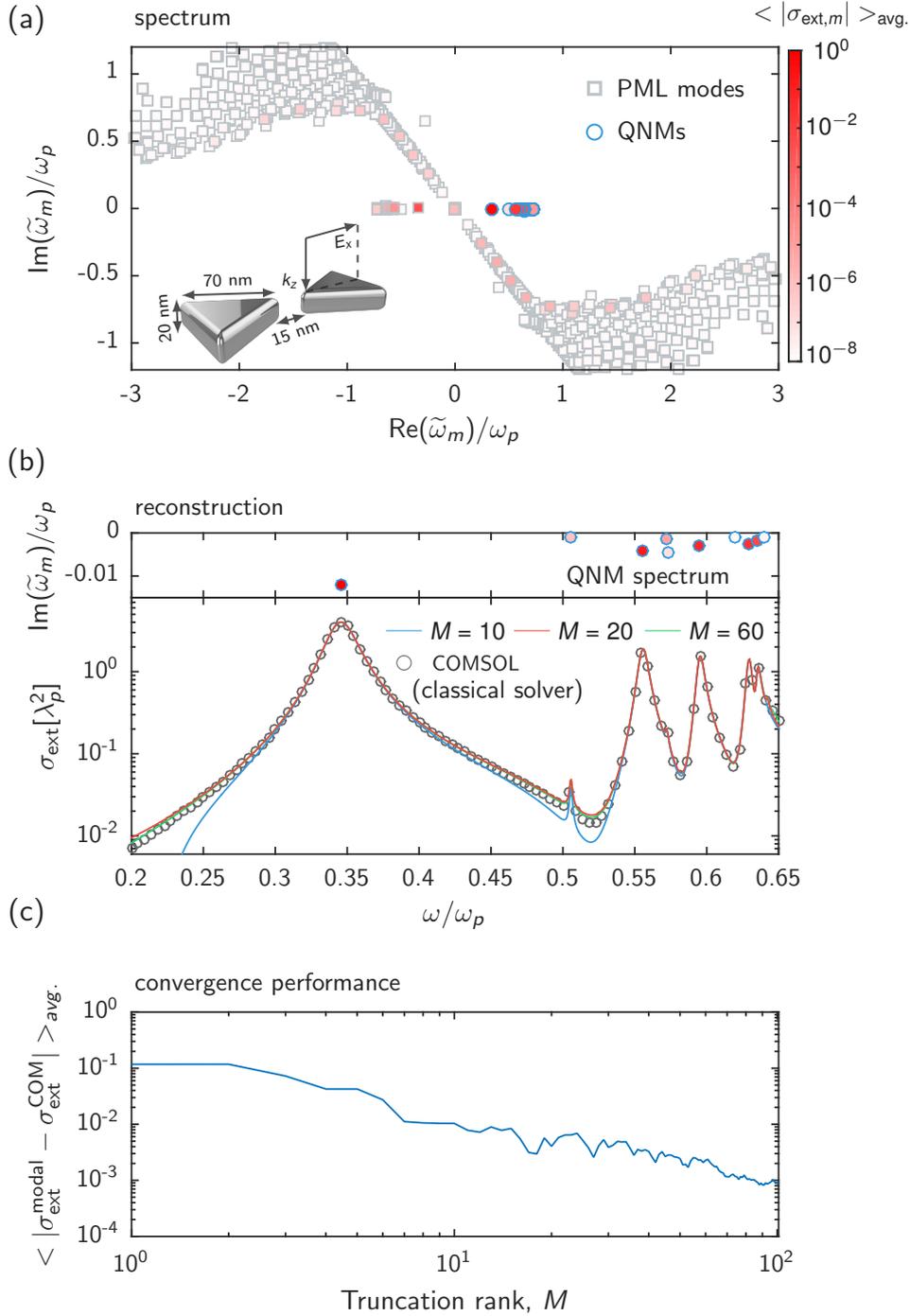

FIG. SI.6 Numerical performance of the modal-expansion method for a silver bowtie in air. The resonator is the same as in Fig. 1 in the main text. (a) Eigenfrequencies of QNMs (circles) and PML-modes (squares) computed with the QNM solver. The color of the markers visualizes $<|\sigma_{\text{ext},m}|>_{\text{avg.}}$, i.e., the contribution of each mode to the spectrally-averaged extinction cross-section. (b) Extinction cross-section spectrum computed with the modal method for $M = 10, 20, 60$ modes and the frequency-domain electromagnetic solver of COMSOL Multiphysics (gray circles). For $M = 10$, we include the 10 dominant QNMs, whose resonance frequencies (see top panel) dictate the position and the width of the resonance peaks. (c) Convergence rate of the modal method. The modes are sorted by increasing order of their impact on the reconstruction.

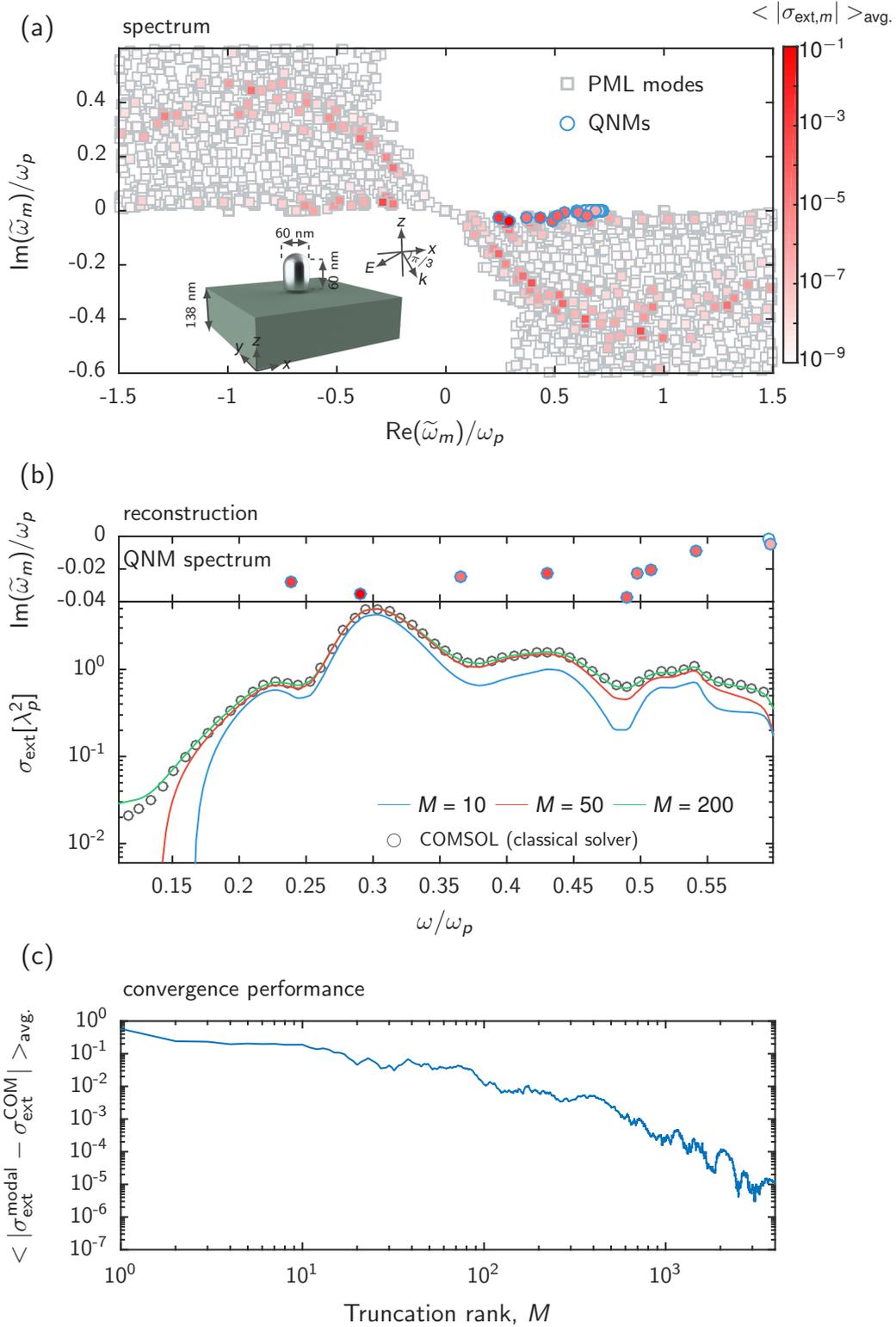

FIG. SI.7 Numerical performance of the modal-expansion method for a silver nanobullet on a semiconductor slab. The resonator is the same as in Fig. 4 in the main text. (a) Eigenfrequencies of QNMs (circles) and PML-modes (squares) computed with the QNM solver. The color of the markers visualizes $<|\sigma_{\text{ext},m}|>_{\text{avg.}}$, i.e., the contribution of each mode to the spectrally-averaged extinction cross-section. (b) Extinction cross-section spectrum computed with the modal method for $M = 10, 50, 200$ modes and the frequency-domain electromagnetic solver of COMSOL Multiphysics (gray circles). For $M = 10$, we include the 10 dominant QNMs, whose resonance frequencies (see top panel) dictate the position and the width of the resonance peaks. (c) Convergence rate of the modal method. The modes are sorted by increasing order of their impact on the reconstruction.

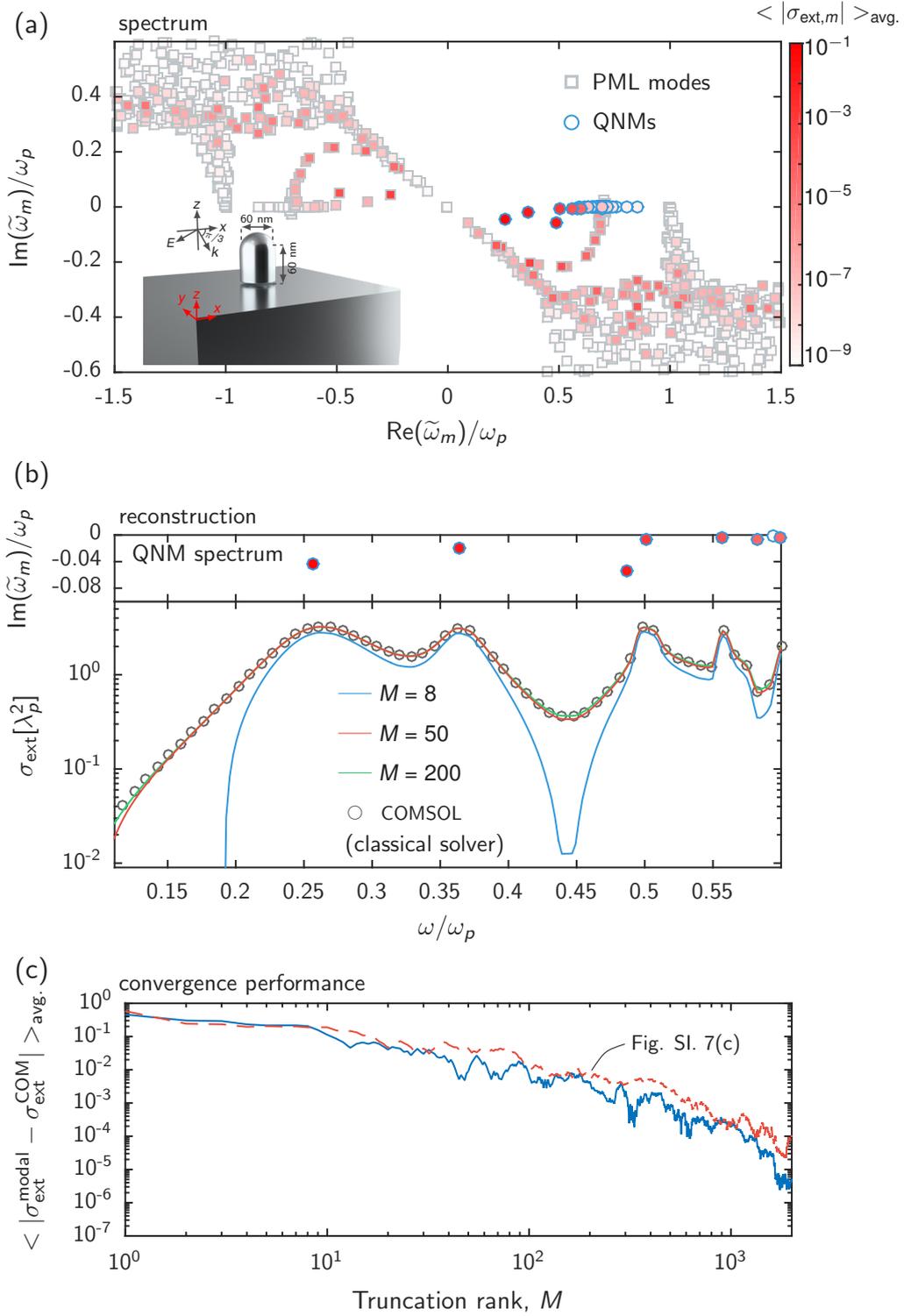

FIG. SI.8 Numerical performance of the modal-expansion method for a silver nanobullet on a silver substrate. The nanobullet is the same as in Fig. SI.7. (a) Eigenfrequencies of QNMs (circles) and PML-modes (squares) computed with the QNM solver. The color of the markers visualizes $<|\sigma_{\text{ext},m}|>_{\text{avg.}}$, i.e., the contribution of each mode to the spectrally-averaged extinction cross-section. (b) Extinction cross-section spectrum computed with the modal method for $M = 8, 50, 200$ modes and the frequency-domain electromagnetic solver of COMSOL Multiphysics (gray circles). For $M = 8$, we includes the 8 dominant QNMs, whose resonance frequencies (see top panel) dictates the position and the width of the resonance peaks. (c) Convergence rate of the modal method. The modes are sorted by increasing order of their impact on the reconstruction.

**5.5. Computational speed**

The goal of this Section is not to compare the computational speed of the modal expansion method with other classical methods for analyzing nanoresonators operating either in the frequency or time domains. Such comparisons are always difficult to make fairly, and additionally, since the present method is still in its infancy, it would be quite premature. Rather, we intend to provide good overall indicators of the computation performance of the software for its present conditions.

Table II gives the CPU times recorded for the computations of the bowtie cross-section spectrum on a PC computer equipped with 3.50 GHz ×2 processors and a 64 GB memory and with Matlab and COMSOL Multphysics 5.0. The CPU times of the QNM computations are mainly determined by the number of degrees of freedom that is proportional to the number of mesh elements and the number of eigenmodes $M$ to be computed. Since the COMSOL eigensolver uses an iterative method to compute the eigenmodes, the CPU time implicitly depends on an iterative number (second column) that usually depends on the mesh and material setting, increases with $M$ generally, and varies with the trial frequency used by the solver to start the initial eigenmode computation.

**Modal expansion**

| Present QNM-eigensolver | | | Cross-section spectra |
|---|---|---|---|
| # modes, $M$ | Iterative number | CPU times | CPU times (200 freq. points) |
| 4 | 3 | 2 mins 13 s | 7 s |
| 12 | 25 | 6 mins 59 s | 17 s |
| 24 | 15 | 9 mins 10 s | 27 s |
| 50 | 20 | 22 mins 50 s | 55 s |

**Frequency-domain FEM (COMSOL Multiphysics)**

| Cross-section spectra |
|---|
| CPU times (200 freq. points) |
| 2 hours 54 mins |

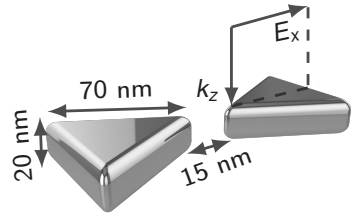

TABLE II  Typical CPU times for computing cross-section spectra. The upper Table refers to the QNM-expansion method for several values of the number of computed QNMs $M = 4, 12, 24,$ and $50$, while the lower Table refers to CPU times observed with the classical frequency-domain FEM solver of COMSOL Multiphysics. The number of degrees of freedom (which scales linearly with matrix size), is $2.76 \times 10^5$ for the frequency-domain FEM solver. It is slightly larger $2.96 \times 10^5$ with the same mesh for the QNM expansion method, because of additional auxiliary fields.

From Table II, several observations can be made:

a. The CPU time is dominantly due to the QNM computations, which is much longer than the cross-section spectrum computation, as shown by the comparison of the third and fourth columns in the up sub-table. The CPU times of the QNM eigensolver approximately scales linearly with the number of computed eigenmodes.

b. The average CPU time per eigenmode computation is approximately 30 seconds. It is twice smaller than the CPU time needed to compute the cross-section at a single frequency with the classical frequency-domain FEM solver (see the lower Table).

c. The computation speed of the modal method is determined by how many modes that we ask the solver to solve. Depending on different problems illustrated in this article, the number of the modes, which are needed for achieving a good numerical accuracy, vary from tens to hundreds. If less modes are needed, the computational advantage of the modal method is obvious. For instance, considering the bow-tie nanoantenna studied in this article, we see that around 10 QNMs can achieve a good numerical accuracy with the overall CPU time around 7 mins and 16s, to be compared with 2 hours and 54 mins for the classical frequency-domain FEM solver if we sample 200 frequency points in the spectrum.

d. The speed advantage of the modal formalism can even be more prominent when more parameters are swept in simulations, e.g., sampling more frequency points or varying polarizations and angles of incident waves. However, we should mention that, when more eigenstates need to be considered in the expansion for accuracy, the gain in speed is lowered. In the time domain, the CPU time for computing the response to a driving pulse is almost the same as the CPU time for computing the cross-section spectra. For the curves shown in Fig. 2 in the main text, it is 7 mins and 20s, a value much smaller than that required with the FDTD method for instance.

**References**

[1] A. Raman and S. Fan, Phy. Rev. Lett. **104**, 087401 (2010).
[2] J. Jackson, *Classical electromagnetics* (John Wiley, New York, 1999).
[3] G. Hanson and A. Yakovlev, *Operator Theory for Electromagnetics: An Introduction* (Springer, 2002).
[4] B. Vial, A. Nicolet, F. Zolla, and M. Commandré, Phys. Rev. A **89**, 023829 (2014).
[5] P. T. Leung, S. Y. Liu, , and K. Young, Phys. Rev. A **49**, 3057 (1994).
[6] P. T. Leung, S. Y. Liu, , and K. Young, Phys. Rev. A **49**, 3982 (1994).
[7] K. M. Lee, P. T. Leung, and K. M. Pang, J. Opt. Soc. Am. B **16**, 14091417 (1999).
[8] L. Novotny and B. Hecht, *Principles of Nano-Optics* (Cambridge University Press, New York, 2006).
[9] C. Sauvan, J.-P. Hugonin, I. S. Maksymov, and P. Lalanne, Phys. Rev. Lett. **110**, 237401 (2013).
[10] E. A. Muljarov and W. Langbein, Phys. Rev. B **94**, 235438 (2016).
[11] R. Ge and S. Hughes, Opt. Lett. **39**, 4235 (2014).
[12] J. Yang, J.-P. Hugonin, and P. Lalanne, ACS Nano **3**, 395 (2016).
[13] C. A. Balanis, *Antenna theory analysis and design* (Wiley-Interscience, 3rd edition, New York, 2005).
[14] K. Demarest, Z. Huang, and R. Plumb, IEEE Trans. on Ant. and Prop. **44**, 1150 (1996).
[15] J. Zimmerling, L. Wei, P. Urbach, and R. Remis, J. Comput. Phys. **315**, 348 (2016).
[16] J. Zimmerling, L. Wei, P. Urbach, and R. Remis, Appl. Phys. A **122**, 158 (2016).
[17] F. Tisseur and K. Meerbergen, SIAM Review **43**, 235 (2001).
[18] A. Taflove, S. Johnson, and A. Oskooi, *Advances in FDTD Computational Electrodynamics: Photonics and Nanotechnology* (Artech Housey, Boston, 2013).
[19] Q. Bai, M. Perrin, C. Sauvan, J. Hugonin, and P. Lalanne, Opt. Express **21**, 27371 (2013).